\documentclass[aps,showpacs,showkeys]{revtex4}
\usepackage{graphicx}
\usepackage{subfigure}
\usepackage{caption2}
\usepackage{float}
\usepackage{amsmath}
\usepackage{amssymb}

\begin{document}

\title{Families of spatial solitons in a two-channel waveguide with the
cubic-quintic nonlinearity}
\author{Ze'ev Birnbaum and Boris A. Malomed}
\affiliation{Department of Physical Electronics, School of Electrical Engineering,
Faculty of Engineering, Tel Aviv University, Tel Aviv 69978, Israel}

\begin{abstract}
We present eight types of spatial optical solitons which are possible in a
model of a planar waveguide that includes a dual-channel trapping structure
and competing (cubic-quintic) nonlinearity. Among the families of trapped
beams are symmetric and antisymmetric solitons of ``broad" and ``narrow"
types, composite states, built as combinations of broad and narrow beams
with identical or opposite signs (``unipolar" and ``bipolar" states,
respectively), and ``single-sided" broad and narrow beams trapped,
essentially, in a single channel. The stability of the families is
investigated via eigenvalues of small perturbations, and is verified in
direct simulations. Three species -- narrow symmetric, broad antisymmetric,
and unipolar composite states -- are unstable to perturbations with real
eigenvalues, while the other five families are stable. The unstable states
do not decay, but, instead, spontaneously transform themselves into
persistent breathers, which, in some cases, demonstrate dynamical symmetry
breaking and chaotic internal oscillations. A noteworthy feature is a
stability exchange between the broad and narrow antisymmetric states: in the
limit when the two channels merge into one, the former species becomes
stable, while the latter one loses its stability. Different branches of the
stationary states are linked by four bifurcations, which take different
forms in the model with the strong and weak inter-channel coupling.
\end{abstract}

\maketitle

\section{Introduction and the model}

Self-trapping of light beams in the form of spatial solitons in planar
waveguides, which was demonstrated experimentally about two decades ago \cite%
{spatial-soliton}, is one of fundamental effects in nonlinear optics . The
variety of the spatial solitons may be greatly expanded if the waveguide is
equipped with a multi-channel structure \cite{Wang}, that can be created by
a permanent transverse modulation of the refractive index, or induced, in a
photorefractive crystal, by a pair of laser beams (with the ordinary
polarization) illuminating the crystal in transverse directions, while the
probe beam, that creates the soliton(s), is launched in the extraordinary
polarization \cite{Moti}. Besides their interest to optical physics,
manipulations of spatial solitons by means of multi-channel trapping
structures have a vast potential for applications to routing of data streams
\cite{Wang}.

The transmission of the electromagnetic wave with local amplitude $\psi
(z,x) $ along axis $z$ in the multi-channel planar waveguide is modeled by
the nonlinear Schr\"{o}dinger (NLS) equation. In the normalized form, the
equation is
\begin{equation}
i\psi _{z}+\psi _{xx}-V(x)\psi +\delta n\left( |\psi |^{2}\right) \psi =0,
\label{NLS}
\end{equation}%
where the second term accounts for the transverse diffraction, $V(x)$
represents the transverse modulation of the refractive index that induces
the multi-channel structure, and term $\sim \delta n(|\psi |^{2})$
represents the optical nonlinearity, $\delta n\left( |\psi |^{2}\right)
=n_{2}|\psi |^{2}$, with $n_{2}>0$, corresponding to the ordinary Kerr
effect. In the model including the Kerr term and simplest transverse
modulation, $V(x)=V_{0}\cos \left( 2\pi x/L\right) $, solitons trapped in
channels and a possibility of switching them between the channels were
studied in Ref. \cite{Wang}. It was found that the entire family of
fundamental (single-humped) solitons, with any value of integral power, $%
P=\int_{-\infty }^{+\infty }|\psi (x)|^{2}dx$, is stable. A family of stable
double-humped bound states of fundamental solitons exists in this model too,
provided that the power exceeds a certain threshold value which grows with
the increase of $V_{0}$. A possibility of the switching, realized as
carrying the fundamental soliton over into an adjacent channel, was also
shown in Ref. \cite{Wang}, under the action of a local kick (``hot spot"),
represented by an additional term $\sim \delta (x-x_{0})\delta (z-z_{0})\psi
$ in Eq. (\ref{NLS}), where $x_{0}$ is a midpoint between the channels.

The same model as the one introduced in Ref. \cite{Wang}, but with $z$
replaced by time $t$ was later considered as an effectively one-dimensional
Gross-Pitaevskii equation for the Bose-Einstein condensate\ (BEC) trapped in
the optical-lattice potential \cite{KonotopEPL} (the general topic of
condensates trapped in optical lattices was reviewed in Refs. \cite%
{BEC-OL-reviews}). In the model with the sinusoidal transverse modulation
replaced by that of the Kronig-Penney (KP) type, i.e., a periodic array of
rectangular potential troughs, soliton states were studied in Ref. \cite%
{Smerzi}, and an effective nonlinear band structure in the same model was
analyzed in detail in Refs. \cite{Carr}. The KP structure is quite relevant
for applications to optics, where it corresponds to the simplest step-index
profile of the transverse modulation.

In addition to employing the multichannel settings, the variety of
spatial-soliton states can be greatly expanded by using media with competing
self-focusing and self-defocusing nonlinearities. A well-known example of
that is provided by the cubic-quintic (CQ) nonlinearity. Optical
nonlinearities of the CQ type were observed in chalcogenide glasses \cite%
{glass}, and in some organic materials \cite{organic}. Actually, the CQ
nonlinear response of these media is induced by an intrinsic resonance,
which also gives rise to nonlinear (two-photon) absorption \cite{Leblond}.
Nevertheless, according to analysis reported in Ref. \cite{YiFan}, effects
of the loss may be neglected in physically relevant settings, as experiments
in optical crystals are conducted over sufficiently short propagation
distances (a few centimeters) \cite{spatial-soliton,Moti}. It is also
relevant to mention that the CQ nonlinearity was predicted \cite{Agarwal}
and recently observed \cite{Brazil} in composite optical media (colloids).

Equation (\ref{NLS}) with the CQ nonlinearity can be cast into the following
normalized form \cite{we,we-all}:
\begin{equation}
i\psi _{z}+\psi _{xx}=V(x)\psi -2|\psi |^{2}\psi +|\psi |^{4}\psi .
\label{eq1}
\end{equation}%
In the uniform medium, which corresponds to $V(x)=0$, Eq. (\ref{eq1}) gives
rise to the well-known family of stable solitons \cite{Canada},
\begin{equation}
\psi _{\mathrm{sol}}(x,z)=e^{ikz}\sqrt{\frac{2k}{1+\sqrt{1-4k/3}\cosh \left(
2\sqrt{k}x\right) }},  \label{soliton}
\end{equation}%
which is parameterized by propagation constant $k$ which takes values in
interval
\begin{equation}
0<k<k_{\max }\equiv 3/4.  \label{interval}
\end{equation}%
In an adjacent interval, $3/4\leq k\leq 1$, solutions exist too, in the form
of continuous-wave (CW) states with a constant amplitude,%
\begin{equation}
\psi _{\mathrm{CW}}^{(\pm )}(z)=e^{ikz}\sqrt{1\pm \sqrt{1-k}}  \label{CW}
\end{equation}%
(the CW solutions exist for all $k\leq 1$). The total power of soliton (\ref%
{soliton}) is%
\begin{equation}
P_{\mathrm{sol}}(k)=\frac{\sqrt{3}}{2}\ln \left( \frac{\sqrt{3}+2\sqrt{k}}{%
\sqrt{3}-2\sqrt{k}}\right) .  \label{P}
\end{equation}

CW solution (\ref{CW}) with the larger amplitude, $\psi _{\mathrm{CW}}^{(+)}$%
, are stable, while solutions $\psi _{\mathrm{CW}}^{(-)}$ are subject to the
modulational instability. At $k=1$, both branches of the CW solutions merge
and disappear through the ordinary tangent bifurcation \cite{JosephIooss}.
Note that the width of soliton (\ref{soliton}) diverges as%
\begin{equation}
W\approx \left( 1/\sqrt{3}\right) \left\vert \ln \left( k_{\max }-k\right)
\right\vert  \label{W}
\end{equation}%
in the limit of $k\rightarrow k_{\max }=3/4$, see Eq. (\ref{interval}), and,
accordingly, the soliton asymptotically goes over into the stable CW state, $%
\psi _{\mathrm{CW}}^{(+)}$, in this limit. In terms of the bifurcation
theory, the disappearance of the bright soliton at this point is explained
by its merger with a dark-soliton solution to Eq. (\ref{eq1}) with $V=0$,
which also exists at $k<k_{\max }\equiv 3/4$. Exactly at $k=3/4$, the dark
soliton degenerates into a \textit{front solution}, which vanishes in one
limit, at $x\rightarrow \pm \infty $, and asymptotically coincides with the
CW state in the other,%
\begin{equation}
\psi _{\mathrm{front}}(x,z)=e^{\left( 3/4\right) iz}\sqrt{\frac{3/2}{%
1+e^{\pm \sqrt{3}x}}}~.  \label{front}
\end{equation}

A natural possibility, which is promising both for the exploration of
fundamental properties of guided solitons and for potential applications, is
to study localized states generated as a result of the interplay between the
CQ nonlinearity and multichannel waveguiding structures. In Ref. \cite{we},
solitons were investigated in the CQ model including a single guiding
channel of a rectangular shape. A distinctive feature of the channel-trapped
solitons in that model is their \textit{bistability}: while in interval (\ref%
{interval}), which hosts solution family (\ref{soliton}) in the free space,
the channel supports a single soliton state for each $k$, the full existence
region of the solitons includes an additional interval, $3/4<k<\tilde{k}%
_{\max }$ (the value of $\tilde{k}_{\max }$ depends on the depth and width
of the guiding channel), where two different solitons are found for a given
value of $k$, \emph{both} being stable (note that a guiding channel does not
give rise to soliton bistability in the model with the cubic nonlinearity
\cite{gh}).

The combination of the CQ nonlinearity with a periodic array of guiding
channels of the KP type was studied in Ref. \cite{we-all}. In addition to
single-humped (fundamental) solitons, many families of stable multi-humped
(higher-order) solutions, which also feature the bistability, were found in
that model. Independently, a model combining the CQ nonlinearity and a
periodic sinusoidal modulation function was introduced in Ref. \cite{China},
where similar results for soliton families were obtained. Note that both
these versions of the model including the CQ terms and periodic potential
function $V(x)$ generate not only soliton families in the semi-infinite
spectral gap, but also solitons in finite bandgaps, provided that the
potential is strong enough. The finite-gap solitons are stable too, but they
do not feature bistability.

The most relevant setting for applications to the all-optical switching is
based on two parallel guiding channels, rather than a periodic array. The
respective model is based on Eq. (\ref{eq1}) with the effective potential in
the form of
\begin{equation}
V(x)=\left\{
\begin{array}{ll}
0, & |x|~<L/2~\mathrm{and~}|x|~>D+L/2, \\
-V_{0}, & L/2<|x|~<D+L/2,%
\end{array}%
\right.  \label{DC}
\end{equation}%
where $V_{0}$, $D$ and $L$ are, respectively, the depth and width of each
channel, and the thickness of the barrier which separates them. This model
is also interesting in terms of the soliton dynamics, as it opens a way to
study \textit{spontaneous symmetry breaking} (SSB) in spatial solitons
supported by the two-channel configuration. The SSB in settings based on
double potential wells has recently drawn a great deal of interest in the
studies of BEC, where it was realized experimentally \cite{Markus}, and
analyzed theoretically, using, chiefly, finite-mode approximations, that
replace the underlying Gross-Pitaevskii equation by a system of ordinary
differential equations \cite{finite-mode}. A conclusion produced by the
analysis is that double-well systems with the self-attractive cubic
nonlinearity (self-focusing, in terms of optics) give rise to stable
asymmetric states (which is known as ``macroscopic quantum self-trapping")
through destabilization of symmetric states, while the systems with the
self-repulsive (defocusing) nonlinearity feature destabilization of
antisymmetric solutions. The latter mechanism also generates stable
asymmetric states.

Actually, a similar finite-mode analysis of the SSB of CW states in a
nonlinear optical coupler (which, for this purpose, is essentially
tantamount to the double-channel waveguide) was published much earlier in
Ref. \cite{Canberra}. That work reported results for the model with the
self-focusing or defocusing Kerr nonlinearity, as well as for a saturable
nonlinearity. SSB in a double-well optical setting has been implemented in a
photorefractive crystal (which corresponds to the saturable nonlinearity)
\cite{Zhigang}.

The objective of the present work is to find symmetric, antisymmetric and
asymmetric self-trapped states (i.e., spatial solitons, in terms of
nonlinear optics) in the two-channel model with the CQ nonlinearity, based
on Eqs. (\ref{eq1}) and (\ref{DC}), and explore bifurcations linking those
states to each other. An obvious difference from the previous works, both
those relying upon the finite-mode approximation \cite{finite-mode} and
works which analyzed the full two-channel configuration in the combination
with the cubic nonlinearity, such as Ref. \cite{Michal}, is the competition
between the self-focusing and self-defocusing nonlinearities, which
drastically alters various states, their stability, and bifurcations.

Note that dynamical switching of nonlinear localized beams in the same model
as considered here was investigated in a recent work \cite{Radik}, by means
of a variational approximation and direct simulations. The analysis had
revealed four different transmission regimes, depending on the total power
of the beam. However, stationary states were not looked for in that work.

The rest of the paper is organized as follows. In Sec. \ref{sec2},
we present basic types of stationary spatial solitons which are
possible in the model, and analyze their stability in Sec.
\ref{sec3}. Depending on the symmetry and width of the solitons, we
identify eight types of the states, five stable and three unstable
(in the same configuration with the cubic nonlinearity, only three
soliton species exist). The stability of each species is established
in a rigorous form, via the computation of eigenvalues for modes of
small perturbations around the solitons, and verified in direct
simulations. For all unstable solutions, the instability eigenvalues
are real. However, in direct simulations the instability does not
destroy the trapped states, but rather transforms them into robust
breathers, that may feature dynamical SSB, and chaotic intrinsic
oscillations, in some cases. In Sec. \ref{sec4}, we demonstrate
generic families (branches) of different soliton species, and
bifurcations that link them together. The bifurcation diagrams are
essentially different in cases of the strong and weak coupling
between the two channels, i.e., roughly speaking, for small and
large values of separation $L$ between them, see Eq. (\ref{DC}). The
solution branches and bifurcation diagrams are displayed in two
different forms, \textit{viz}., as the total power versus the
propagation constant, cf. Eq. (\ref{P}), and also as an effective
asymmetry parameter of the state (it is defined below as \ per Eq.
(\ref{epsilon})) versus the total power. The paper is concluded by
Sec. \ref{sec5}.

\section{Basic types of spatial solitons in the two-channel system}
\label{sec2}

In the model with the CQ nonlinearity, each waveguiding channel can support
localized states of two different types, which we will call narrow and broad
ones. The ``narrow" state is nothing else but soliton (\ref{soliton})
trapped in the channel, while its ``broad" counterpart may be realized as a
trapped fragment of CW state (\ref{CW}). In fact, the bistability of soliton
states found in the CQ model with the single channel \cite{we} is accounted
by the coexistence of stable solutions of these two types. In the
two-channel system, these states may be combined in different ways. First,
symmetric and antisymmetric spatial solitons, each built of narrow (Fig. \ref%
{fig1}) or broad (Fig. \ref{fig2}) beams trapped in each channel, is easily
found as a numerical solution of the equation obtained from Eq. (\ref{eq1})
by the substitution of $\psi (z,x)=\exp \left( ikz\right) u(x)$ with real
function $u(x)$,%
\begin{equation}
u^{\prime \prime }=ku+V(x)u-2u^{3}+u^{5}.  \label{ODE}
\end{equation}%
Numerical solution of Eq. (\ref{ODE}) was performed by means of the
relaxation method based on the Newton-Raphson algorithm (formally
speaking, this equation with $V(x)$ taken in the form given by Eq.
(\ref{DC}) can be solved analytically, as explicit solutions in
terms of elliptic functions are available in each region where
$V(x)$ takes a constant value; however, the continuity conditions
for $u(x)$ and $u^{\prime }(x)$ at points $|x|=L/2$ and $|x|=D+L/2$
give rise to very cumbersome transcendental equations).
\begin{figure}[tbp]
\subfigure[]{\includegraphics[width=1.75in]{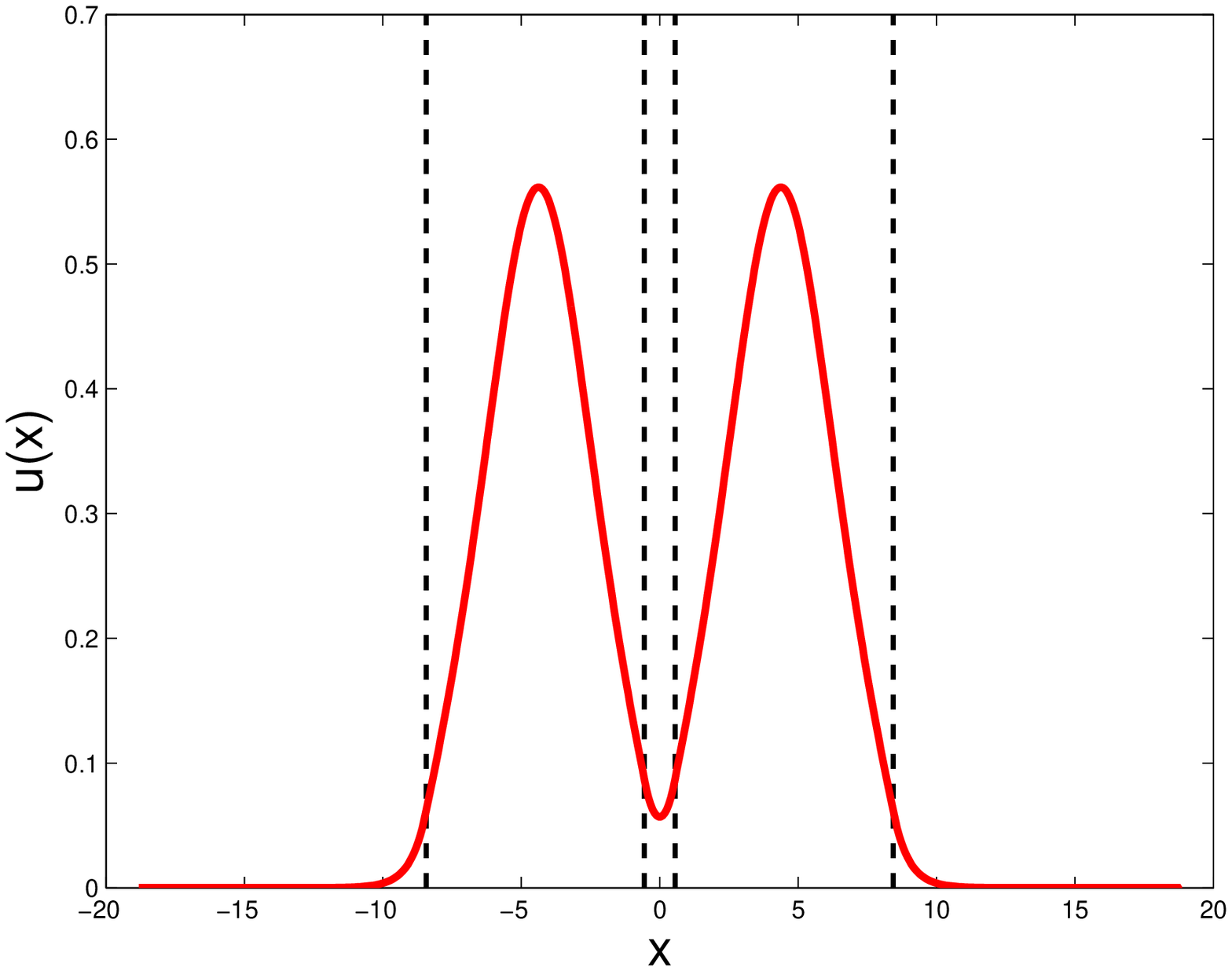}} \subfigure[]{%
\includegraphics[width=1.75in]{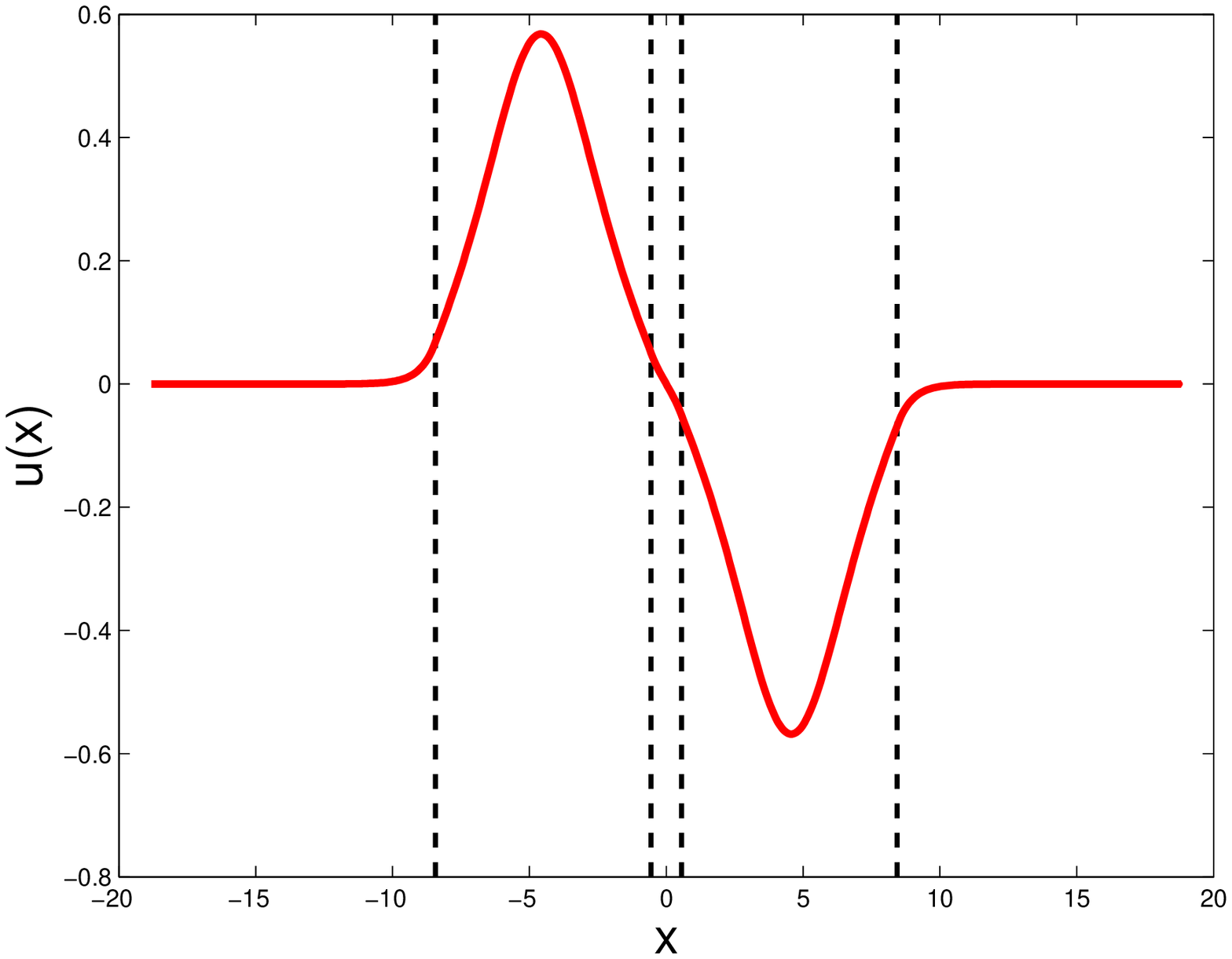}}
\caption{(Color online) Typical examples of symmetric (a) and antisymmetric
(b) spatial solitons built of narrow beams trapped in the two channels.
These examples and those displayed below for other generic types of spatial
solitons in the two-channel system are displayed for the two-channel
potential (\protect\ref{DC}) with $V_{0}=3$, $D=8$, $L=1$, and propagation
constant $k=3.25$, see Eq. (\protect\ref{ODE}). Borders of the trapping
channels are shown (here and in other figures) by vertical dashed lined.}
\label{fig1}
\end{figure}
\begin{figure}[tbp]
\subfigure[]{\includegraphics[width=1.75in]{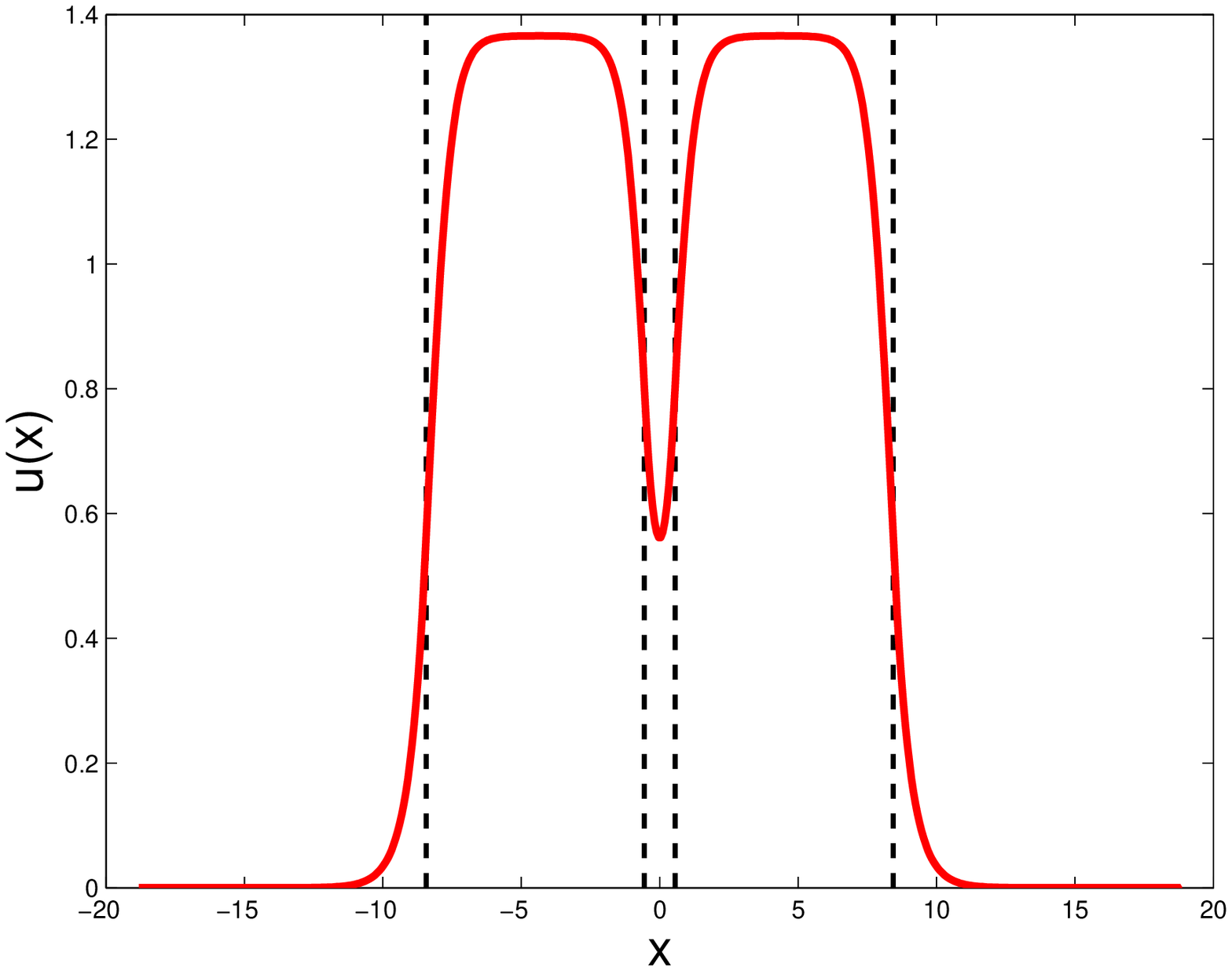}} \subfigure[]{%
\includegraphics[width=1.75in]{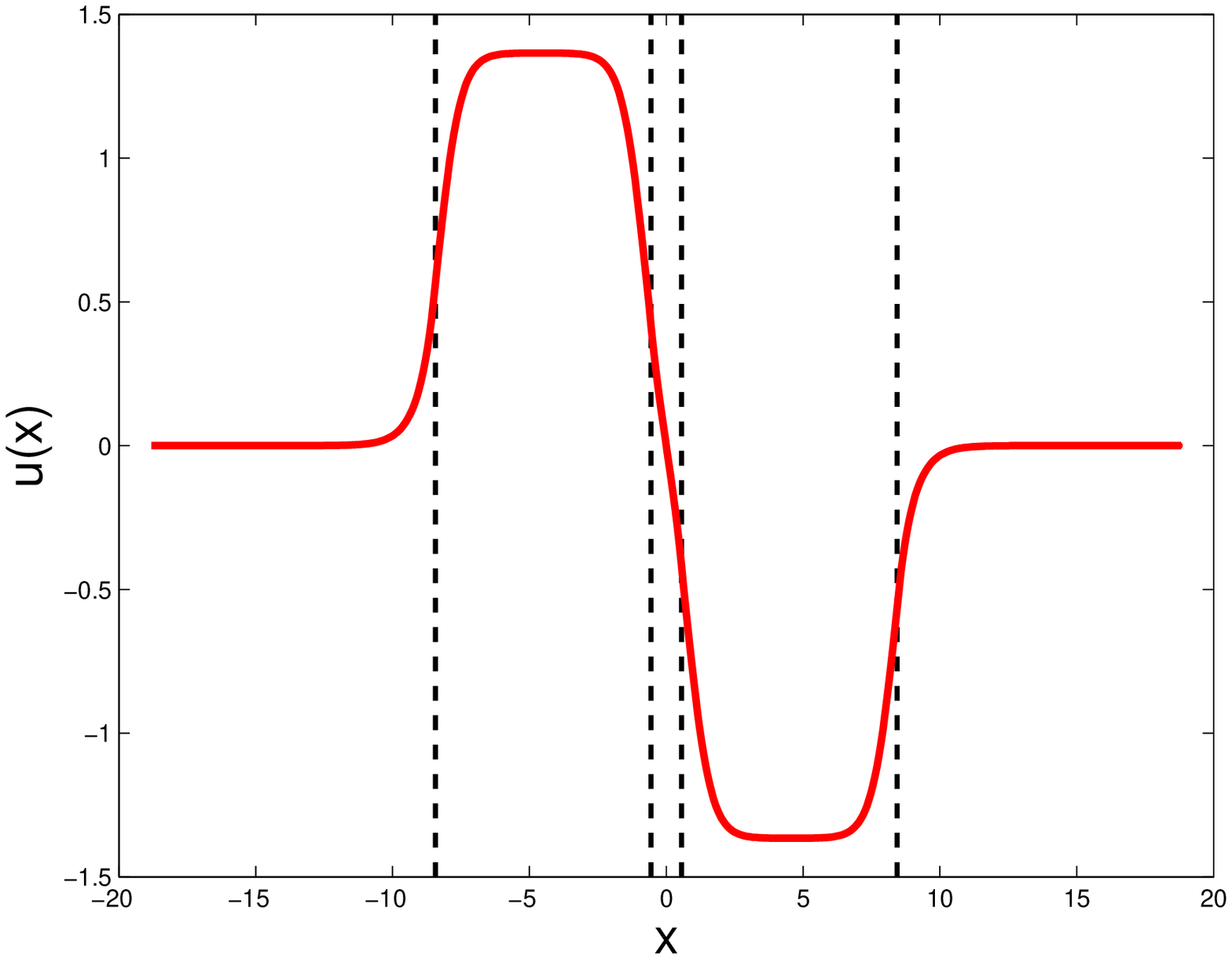}}
\caption{(Color online) Typical examples of symmetric (a) and antisymmetric
(b) spatial solitons formed by broad pulses trapped in the two channels.
Parameters are the same as in Fig. \protect\ref{fig1}.}
\label{fig2}
\end{figure}

As mentioned above, the NLS equation with the CQ nonlinearity in the free
space, i.e., Eq. (\ref{eq1}) with $V(x)=0$, admits not only exact solutions
in the form of the bright solitons and CW states, as given by Eqs. (\ref%
{soliton}) and (\ref{CW}), but also dark-soliton solutions. Accordingly, the
antisymmetric bound states of two broad beams, a generic example of which is
displayed in Fig. \ref{fig2}(b), may be realized as a ``curtailed" dark
soliton (the one whose flat background was chopped off) trapped in the
two-channel setting.

\begin{figure}[tbp]
\subfigure[]{\includegraphics[width=1.75in]{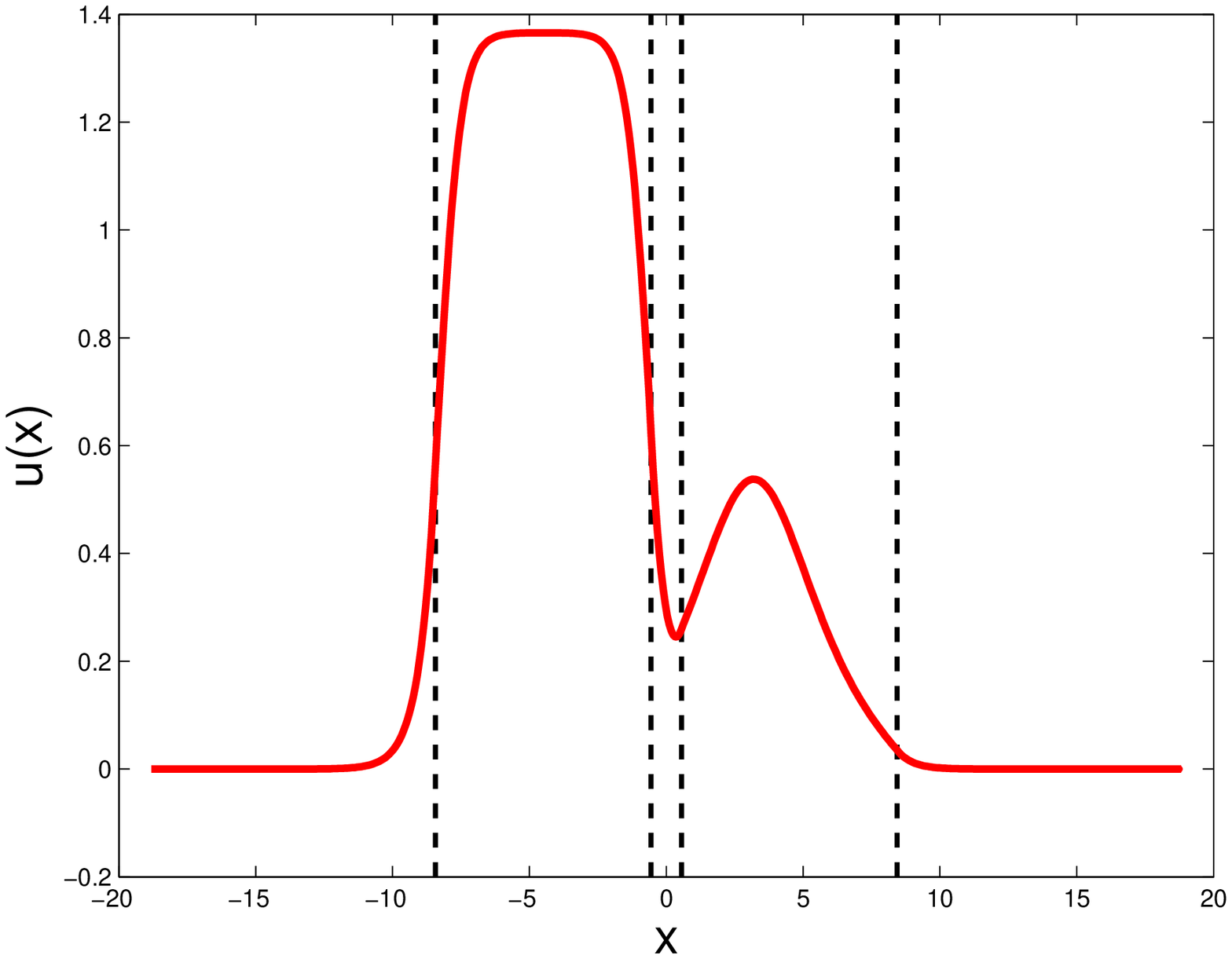}} \subfigure[]{%
\includegraphics[width=1.75in]{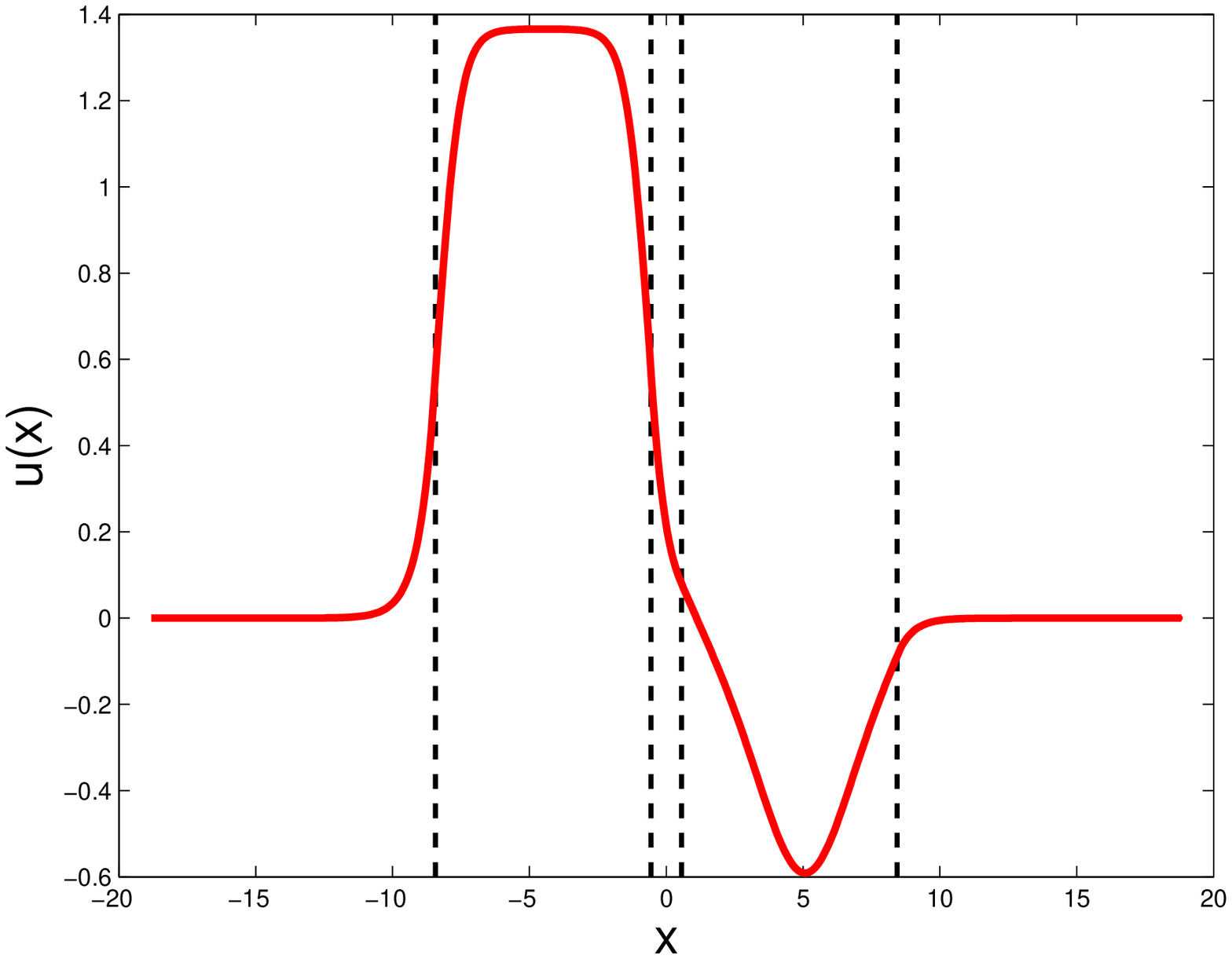}}
\caption{(Color online) Typical examples of unipolar (a) and bipolar (b)
asymmetric spatial solitons built as bound states of broad and narrow pulses
trapped in the two channels. Parameters are the same as in Fig. \protect\ref%
{fig1}.}
\label{fig3}
\end{figure}

Further, using a broad beam placed in one channel, and its narrow
counterpart in the other, it is easy to find solutions to Eq. (\ref{ODE}) in
the form of asymmetric \textit{composite states}, as shown in Fig. \ref{fig3}%
. The composite states, as well as the bistability of spatial solitons
trapped in a single channel \cite{we}, are specific features of the model
with the CQ nonlinearity (one may expect the same features in models with
more general competing nonlinearities), which are not supported by
nonlinearities of \textit{non-competing} types, such as those represented by
cubic or saturable terms in the NLS equation. The signs of the two
components of the composite spatial solitons may be identical or opposite,
see Figs. \ref{fig3}(a) and \ref{fig3}(b). We will refer to these two
species as \textit{unipolar} and \textit{bipolar} composite solitons,
respectively.

\begin{figure}[tbp]
\includegraphics[width=3.0in]{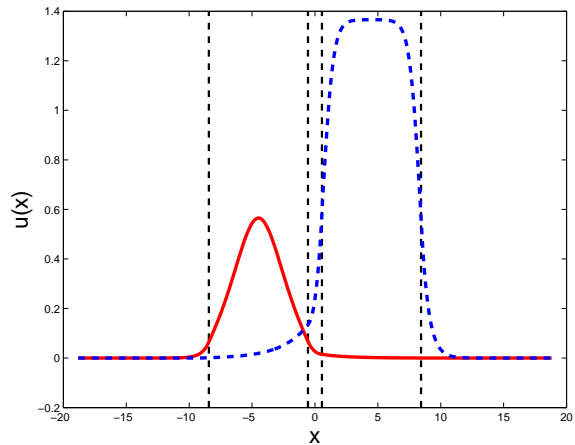}
\caption{(Color online) Generic examples of the single-sided states, with
one channel carrying a narrow or broad pulse (they are shown, respectively,
by solid and dashed curves in the left and right parts of the figure), while
the other channel is nearly empty. Parameters are the same as in Fig.
\protect\ref{fig1}.}
\label{fig4}
\end{figure}

Asymmetric states that, as said above, are well known in models combining
double-well potentials and cubic or saturable nonlinearity \cite{finite-mode}%
-\cite{Michal}, have their straightforward counterparts in the present
model, in the form of a narrow or broad beam trapped in one channel, while
the other channel is left almost empty. Typical examples of these (strongly
asymmetric) states, which are different from the above-mentioned moderately
asymmetric composite spatial solitons, are displayed in Fig. \ref{fig4}.
Below, we refer to them as ``single-sided" states.

\begin{figure}[tbp]
\includegraphics[width=3.0in]{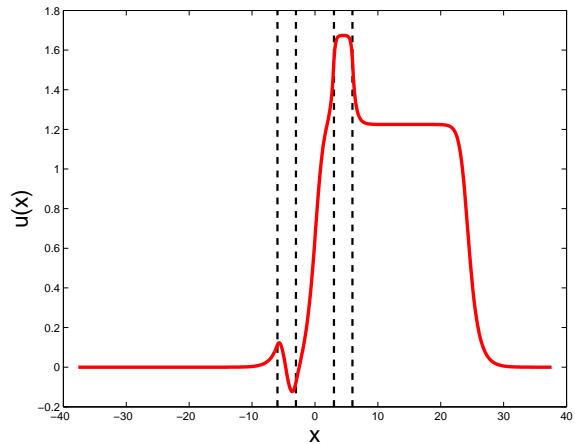}
\caption{(Color online) An example of a stable single-sided broad beam,
which is very close to the transition into a front-shaped state. Parameters
of the two-channel potential are $V_{0}=3$, $D=3$, $L=6$, and the
propagation constant is extremely close to value $k_{\max }=3/4$ which
corresponds to the transition between the dark-soliton and CW states, via
front configuration (\protect\ref{front}), in the free space (see Eq. (%
\protect\ref{interval})), $k=0.75+10^{-16}$.} \label{fig5}
\end{figure}

A noteworthy feature of the single-sided broad-beam solution is that, for $k$
approaching the maximum value, $k_{\max }=3/4$, above which soliton (\ref%
{soliton}) cannot exist in the free space, according to Eqs. (\ref{interval}%
) and (\ref{W}), and is replaced by the spatially infinite CW state, $\psi _{%
\mathrm{CW}}^{(+)}$ (see Eq. (\ref{CW})), the broad-beam solution
goes over not into something close to $\psi _{\mathrm{CW}}^{(+)}$,
but, instead, into a front state which is similar to exact solution
(\ref{front}), as shown in Fig. \ref{fig5}. Recall that, in the free
space, the front solution exists at a single value of the
propagation constant, $k=3/4$, while an external
potential may support a family of front solutions \cite{DW} (in Ref. \cite%
{DW}, fronts were represented by solutions to the Gross-Pitaevskii equation
which included the cubic nonlinearity and a sinusoidal potential). In the
present model, it can be shown that front states may also be sustained by a
single potential well, in the absence of the second channel. We note that
such a solution (which is stable, see below) was not reported in the
previous analysis of the single-channel CQ model \cite{we}.

Examples of the eight different species of spatial solitons possible
in the present model (symmetric and antisymmetric narrow and broad
states, unipolar and bipolar composite ones, and broad or narrow
single-sided solutions) were shown in Figs. \ref{fig1}-\ref{fig4}
for a set of parameters at which the coupling between the channels
is relatively strong (this is seen in the lack
of strong separation between two components of the solitons in Figs. \ref%
{fig1}(a), \ref{fig2}(a), and \ref{fig3}(a)). In the limit of
$L\rightarrow 0 $, when the buffer layer that separates the channels
disappears, and we thus return to the single-channel setting, the
symmetric states continuously go over into single-humped solitons
(narrow or broad ones), which were studied in Ref. \cite{we}. In the
same limit, the antisymmetric states gradually turn into a dipole
(soliton-antisoliton pair) trapped in the
remaining single channel. These transitions are illustrated by Fig. \ref%
{fig6}, which shows the total power of each symmetric and antisymmetric
soliton family, $P$, versus $L$ -- from $L=0$, that corresponds to a single
channel of width $2D $, up to large values of $L$, that correspond to
isolated channels of width $D$. Note that the power of the narrow symmetric
state in the limit of large $L$, i.e., the total power of the set of two
fundamental solitons, is naturally very close to $2P(L=0)$, since the power
at $L=0$ is that of a single soliton. On the other hand, it is natural too
that powers of the broad states trapped in the single channel of the double
width (at $L=0$), and in two widely separated channels (at larger $L$) are
nearly equal, as, for the nearly-CW configurations, $P$ is proportional to
the total width of the area filled by the wave.
\begin{figure}[tbp]
\subfigure[]{\includegraphics[width=1.75in]{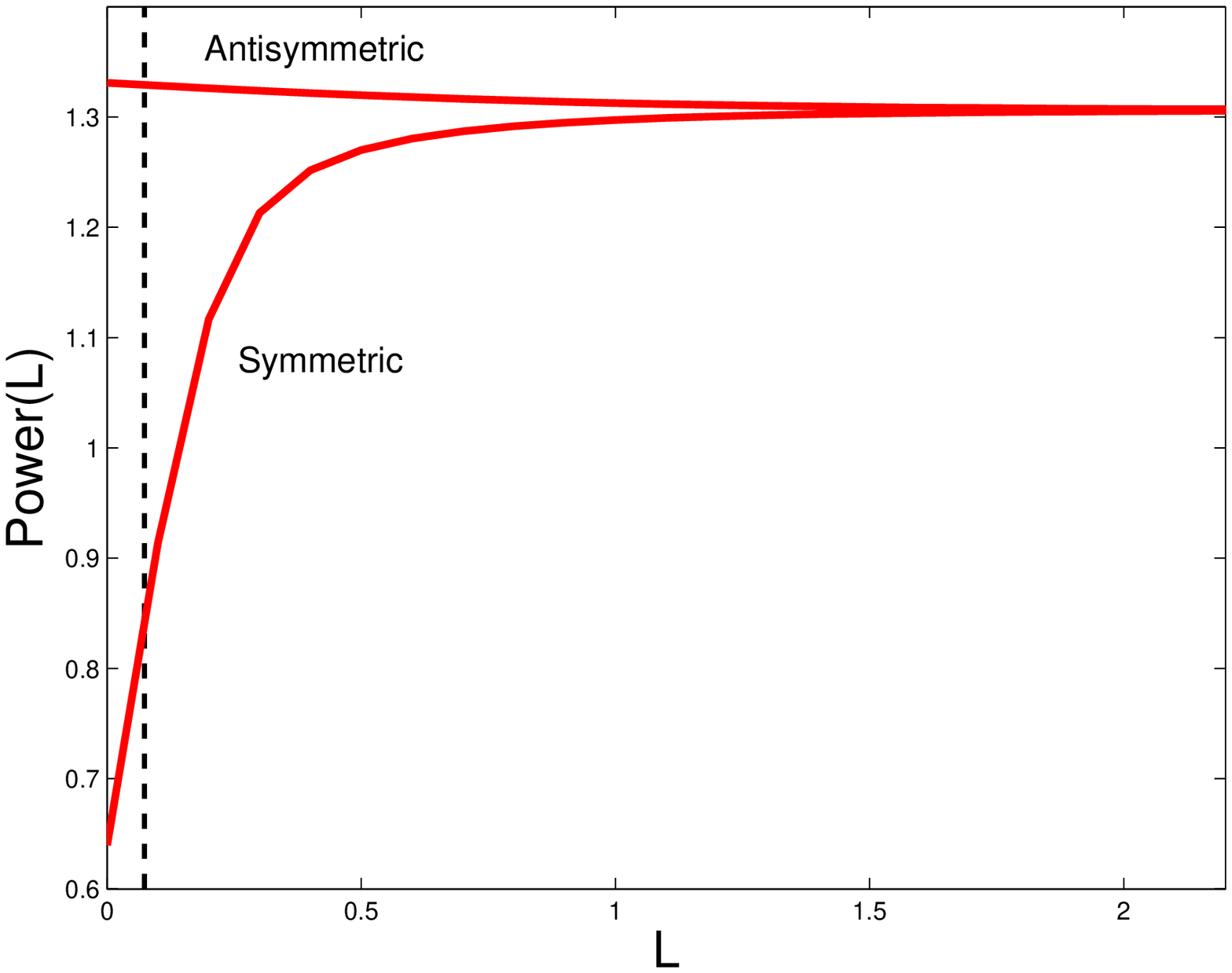}} \subfigure[]{%
\includegraphics[width=1.75in]{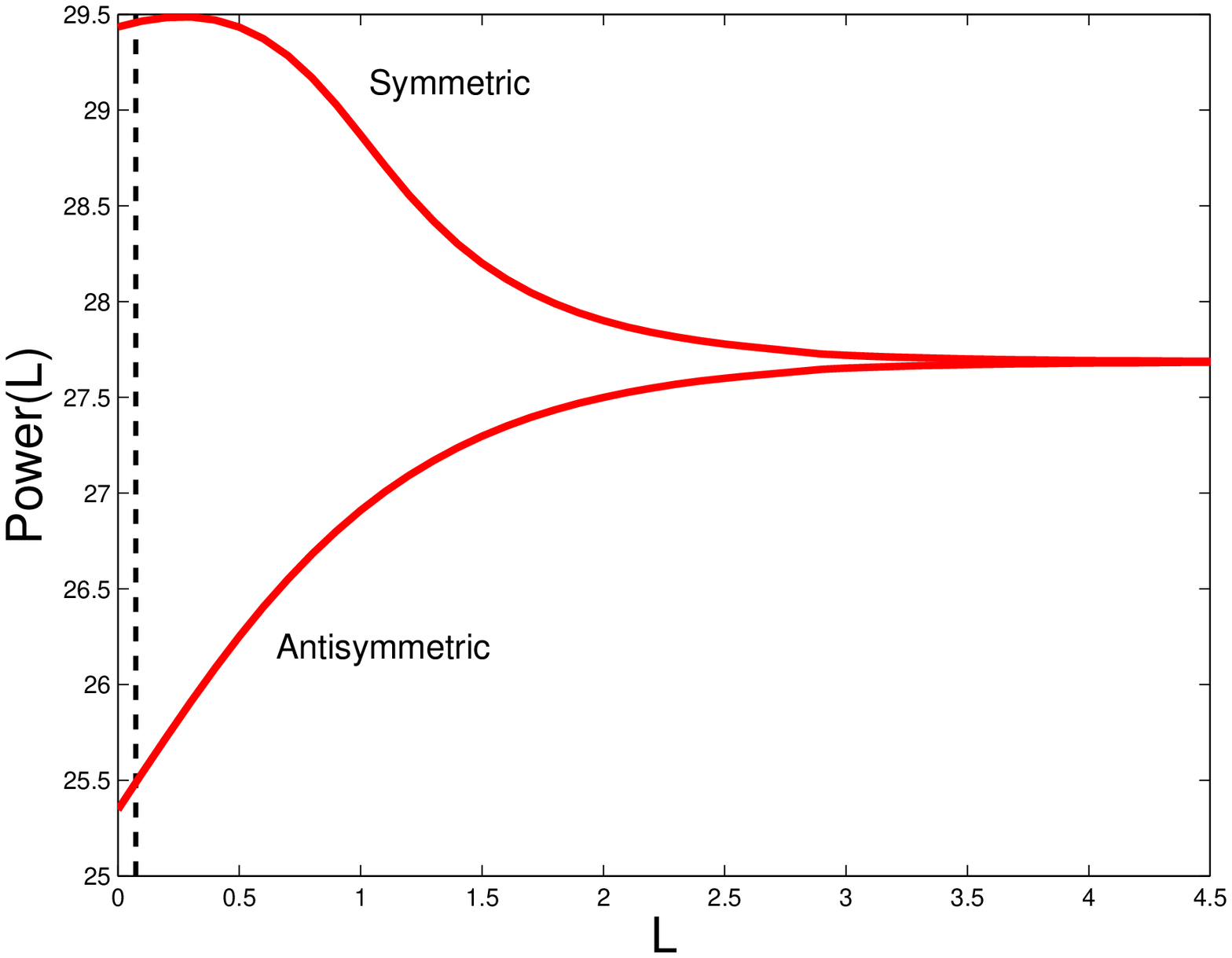}}
\caption{(Color online) The total power of the antisymmetric and symmetric
narrow (a) and broad (b) solitons as a function of thickness $L$ of the
buffer layer between the channels. Fixed parameters are $V_{0}=3$, $D=8$,
and $k=3.1$. The dashed vertical lines mark small values of $L$ at which the
stability is switched, see text.}
\label{fig6}
\end{figure}

Unlike the symmetric and antisymmetric states, the evolution of asymmetric
ones, i.e., the trapped beams of the unipolar and bipolar composite and
single-sided types, with the decrease of $L$ is not continuous. It was found
that, at some small critical value of $L$, each asymmetric state performs a
jump, either into a symmetric state (unipolar composite and single-sided
spatial solitons), or into an antisymmetric one (the bipolar composite
spatial soliton).

\begin{figure}[tbp]
\includegraphics[width=3.0in]{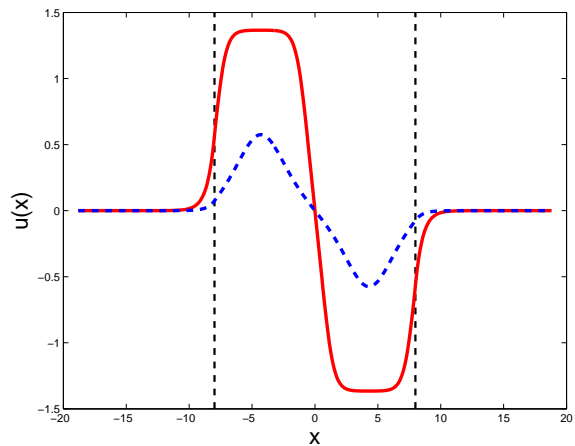}
\caption{(Color online) An example of antisymmetric spatial solitons
(dipoles), narrow and broad ones (shown by the dashed and solid curves,
respectively), trapped in the single channel of width $2D$ (edges of the
channel are marked by dashed vertical lines). Parameters are $V_{0}=3$, $%
2D=16$, and $k=3.1$. The narrow dipole is unstable, while its broad
counterpart is stable.}
\label{fig7}
\end{figure}

Examples of antisymmetric states found in the limit of $L=0$ in the single
channel are displayed in Fig. \ref{fig7}. Such dipole states were not
considered in the framework of the previous work dealing with spatial
solitons in the single-channel CQ model \cite{we}. However, they are
qualitatively similar to the so-called ``dark-in-the-bright" solitons that
were investigated as solutions to the Gross-Pitaevskii equation with a
parabolic confining potential \cite{Yannis}. The peculiarity of the model
with the CQ nonlinearity is that it provides for the coexistence of two
different trapped dipoles at a common value of $k$, similar to the
bistability of the trapped fundamental solitons. A noteworthy finding is
that the stability of narrow and broad antisymmetric solitons trapped in a
single channel is \emph{opposite} to that of their counterparts in the
two-channel setting, where the families of narrow and broad antisymmetric
states are, respectively, stable and unstable (see below): in the single
channel, the narrow dipole is unstable, while the broad one is stable. In
fact, the stability switch occurs at very small finite values of $L$, which
are marked by dashed vertical lines in Fig. \ref{fig6}.

In addition to the states trapped in the channels, we have also found
solutions for beams trapped in the buffer layer between the channels. Such
settings are known as nonlinear antiwaveguides \cite{gh,Gisin}. We do not
display those solutions here because, as may be naturally expected, they are
subject to a strong instability.

\section{Stability of the spatial solitons}
\label{sec3}

Stability of the eight families of spatial solitons found in the two-channel
setting was investigated in two complementary ways, \textit{viz.}, by
numerical computation of stability eigenvalues from linearized equations for
small perturbations, and by means of direct simulations of perturbed
solitons. Both methods have produced identical conclusions concerning the
stability or instability of each family of the stationary states.

For the analysis of small perturbations, the solution was looked for as%
\begin{equation}
\psi (x,z)=e^{ikz}\left[ u(x)+\epsilon \left( U(x)\exp \left( \sigma
z\right) +V^{\ast }(x\right) \exp \left( \sigma ^{\ast }z\right) \right] ,
\label{perturbed}
\end{equation}%
where $u(x)$ is a real stationary solution pertaining to propagation
constant $k$, the asterisk stands for the complex conjugation, $\epsilon $
is a real infinitesimal amplitude of the perturbation with complex
eigenmodes $U(x)$ and $V(x)$, that are associated to a (generally, complex)
eigenvalue $\sigma $. As usual, the stability condition is $\mathrm{Re}%
(\sigma )=0$, which must hold for all the eigenvalues. The substitution of
expression (\ref{perturbed}) in Eq. (\ref{eq1}) and linearization yield the
eigenvalue problem that can be written a matrix form,%
\begin{equation}
\left(
\begin{array}{cc}
\hat{L}+i\sigma & 2u^{2}(x)\left[ 1-u^{2}(x)\right] \\
2u^{2}(x)\left[ 1-u^{2}(x)\right] & \hat{L}-i\sigma%
\end{array}%
\right) \left(
\begin{array}{c}
U \\
V%
\end{array}%
\right) =0,  \label{eigen}
\end{equation}%
where the Sturm-Liouville operator is $\hat{L}\equiv
d^{2}/dx^{2}-V(x)+u^{2}(x)\left[ 4-3u^{2}(x)\right] -k$. Numerical analysis
of the stability spectrum was performed by the computation of eigenvalues of
a large-size matrix approximating the operator on the left-hand side of Eq. (%
\ref{eigen}).

The following results have been obtained. Five species of the spatial
solitons are \emph{stable} in \emph{entire domains} of their existence
(i.e., all the respective stability eigenvalues have zero real parts, up to
the accuracy of the numerical computation), with the exception of the
above-mentioned stability switch of the antisymmetric states at very small
values of $L$. Thus, the following families of stable solutions can be
identified:

(i) Narrow antisymmetric states, an example of which is displayed above in
Fig. \ref{fig1}(b).

(ii) Broad symmetric states, see an example in Fig. \ref{fig2}(a).

(iii) Bipolar composite states, see an example in Fig. \ref{fig3}(b).

(iv,v) Both narrow and broad single-sided states, see Fig. \ref{fig4}.

Other three species, i.e., broad antisymmetric, narrow symmetric, and
unipolar composite states, are unstable, also in their entire existence
domains (except for the broad antisymmetric state at very small values of $L$%
, where it becomes stable, as mentioned above). Eigenvalues accounting for
the instability of these three species are always real. In fact, the
instability growth rate for the broad antisymmetric states (typically, it
takes values $\sigma \sim 0.2$) is usually larger than its counterpart for
the narrow symmetric and unipolar composite states, with the same parameters
of the model, by a factor $\sim 10$.

We note that the instability of the narrow symmetric states and the
stability of their antisymmetric counterparts might be expected, as the
general analysis developed in Refs. \cite{Todd} predicts that bound states
of two solitons supported by an external potential are unstable if the phase
shift between the solitons is $\Delta \phi =0$, and they may be stable if $%
\Delta \phi =\pi $. This prediction does not apply to the symmetric and
antisymmetric broad states, as they originate not from solitons proper, but
rather from trapped segments of CW states, as discussed above.

It is relevant to iterate that the stability of the antisymmetric states is
swapped at very small values of $L$, which implies the transition from the
two-channel setting to the single channel: as shown in Fig. \ref{fig6}, at $%
L\rightarrow 0$ the broad antisymmetric solution becomes stable, while its
narrow counterpart loses its stability. In fact, broad antisymmetric states
trapped in the single wide channel resemble dark solitons, and their
stability complies with the known fact that dark solitons are stable in the
Gross-Pitaevskii equation combining the self-repulsive cubic nonlinearity
and parabolic potential trap \cite{dark-stable}.

Predictions of the analysis based on the computation of the eigenvalues were
checked in direct simulations of the evolution of perturbed spatial
solitons, of all the eight species found above. The simulations of Eq. (\ref%
{eq1}) with the respective initial conditions were performed by means of the
usual split-step method. They identify stable and unstable species of the
solitons in full agreement with the predictions of the linear-stability
analysis. In particular, all the five species of the spatial solitons that
are expected to be stable, listed above under numbers (i) through (v),
indeed demonstrate very robust evolution in the presence of perturbations
(examples of the stable evolution are not displayed here, as they do not
reveal anything essentially new), except for the above-mentioned instability
of the narrow antisymmetric state at $L\rightarrow 0$.

\begin{figure}[tbp]
\subfigure[]{\includegraphics[width=3.0in]{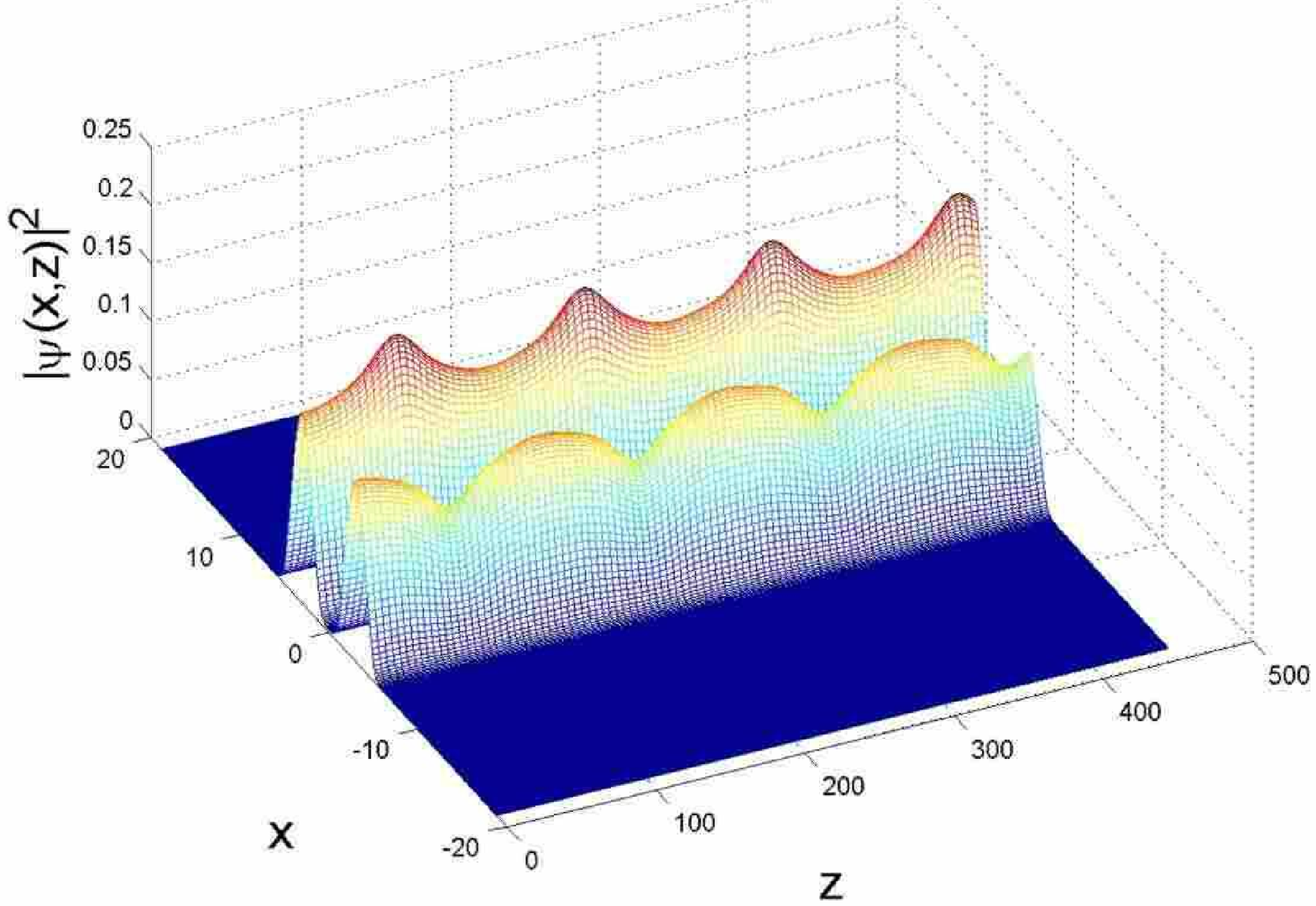}} \subfigure[]{%
\includegraphics[width=1.5in]{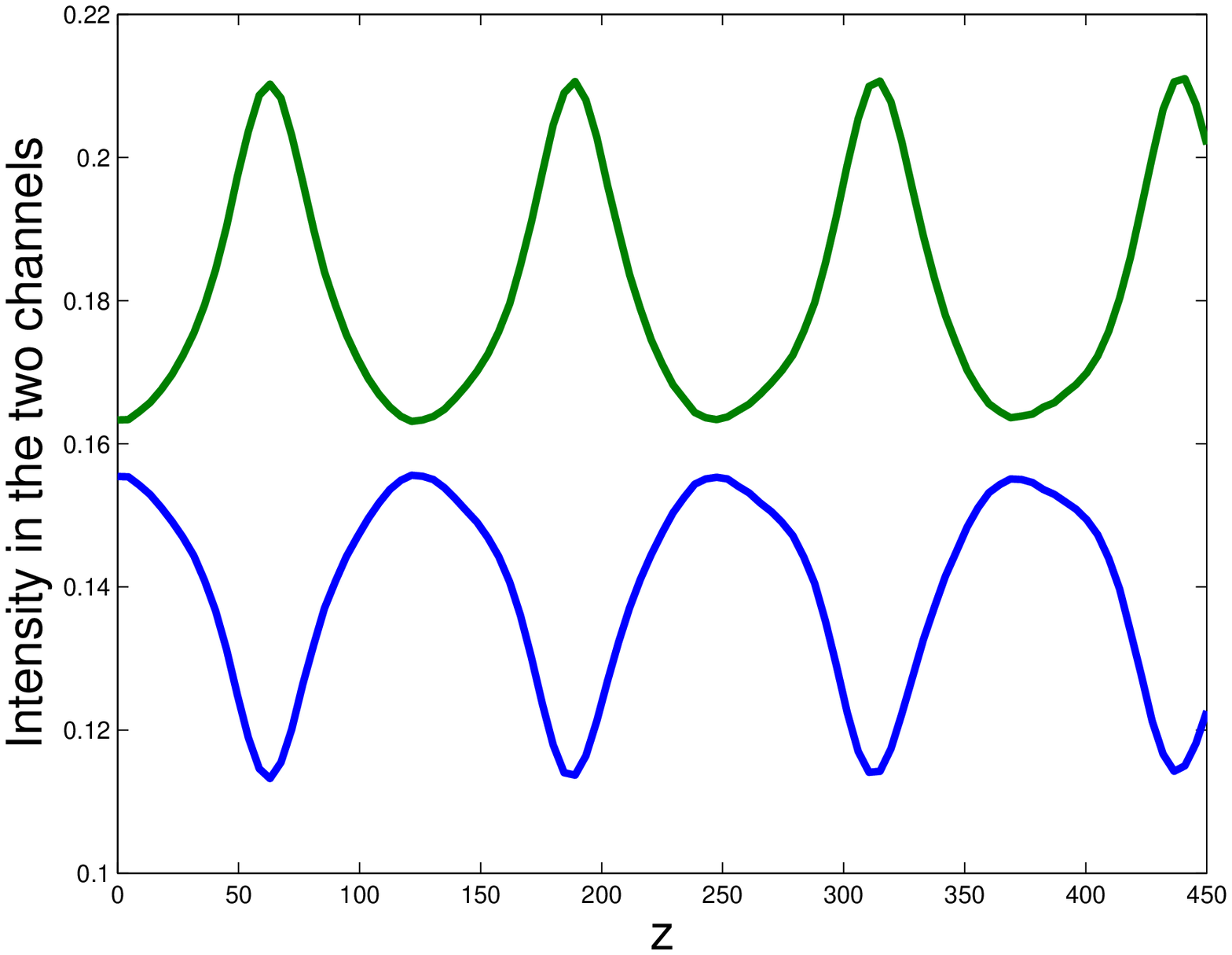}}
\caption{(Color online) Transformation of an unstable narrow symmetric
stationary soliton into a persistent breather, initiated by a weak initial
perturbation: (a) the evolution of the local intentisty; (b) peak
intensities in the two channels versus $z$. Parameters of the two-channel
trapping potential are $V_{0}=3$, $D=4$, $L=2$, and the propagation constant
of the unperturbed state is $k=$ $2.85$. The calculation based on Eq. (%
\protect\ref{eigen}) yields instability growth rate $\protect\sigma \approx
0.05$ for the underlying stationary soliton.}
\label{fig8}
\end{figure}

\begin{figure}[tbp]
\subfigure[]{\includegraphics[width=3.0in]{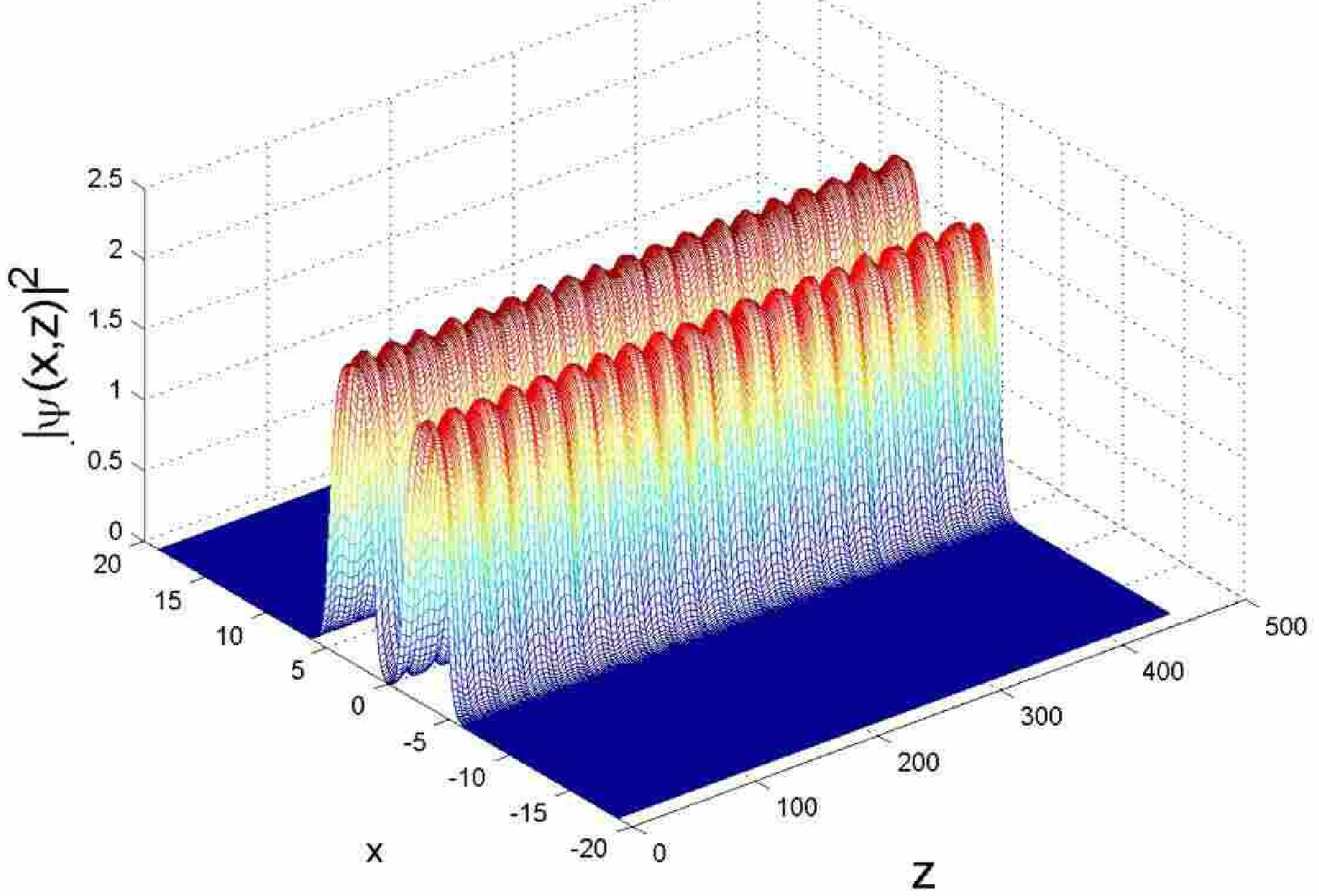}} \subfigure[]{%
\includegraphics[width=1.5in]{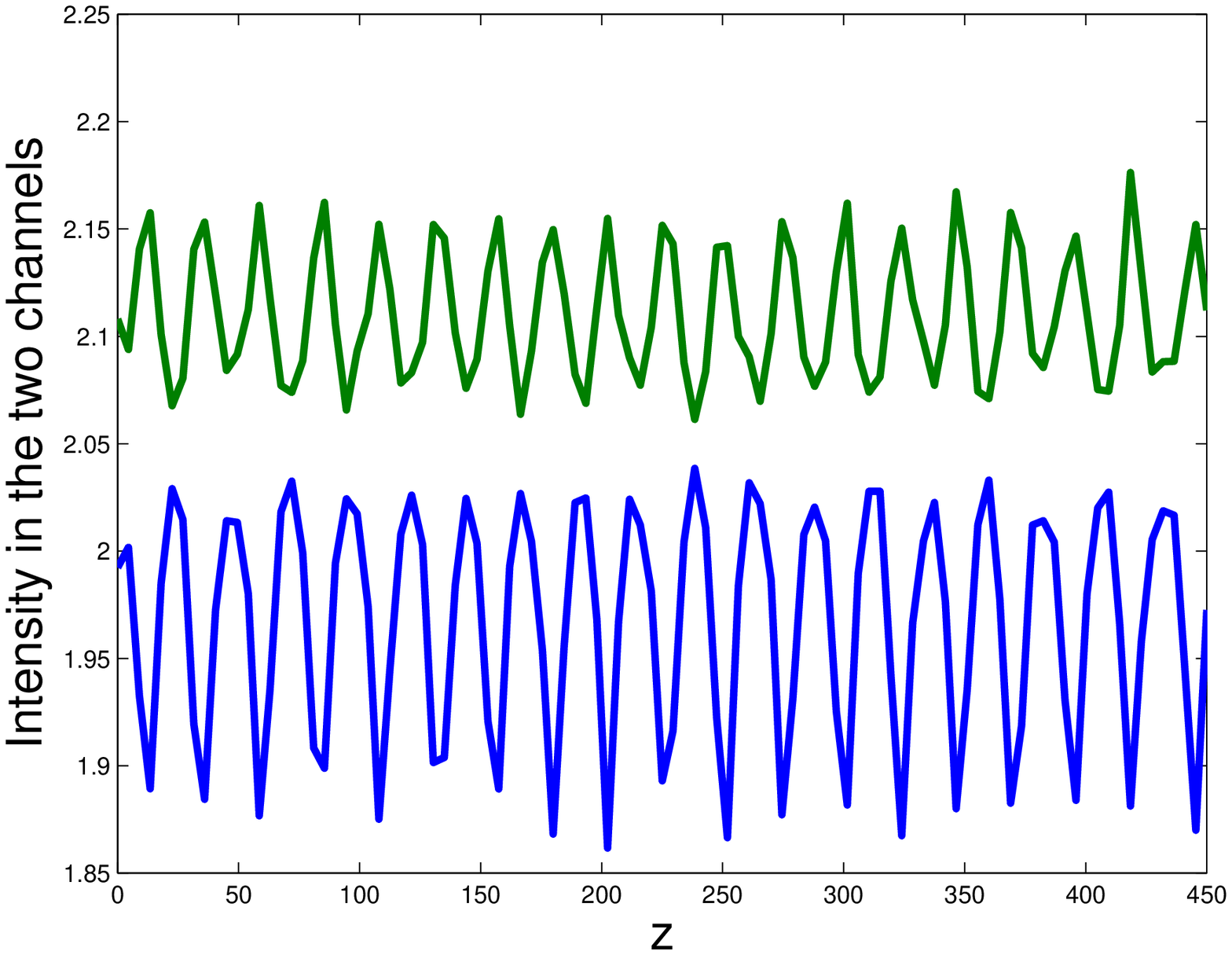}}
\caption{(Color online) The same as in Fig. \protect\ref{fig8} (with the
same values of the parameters) for an unstable broad antisymmetric soliton.
In this case, the instability growth rate for the underlying stationary
soliton is $\protect\sigma \approx 0.25$.}
\label{fig9}
\end{figure}

\begin{figure}[tbp]
\subfigure[]{\includegraphics[width=3.0in]{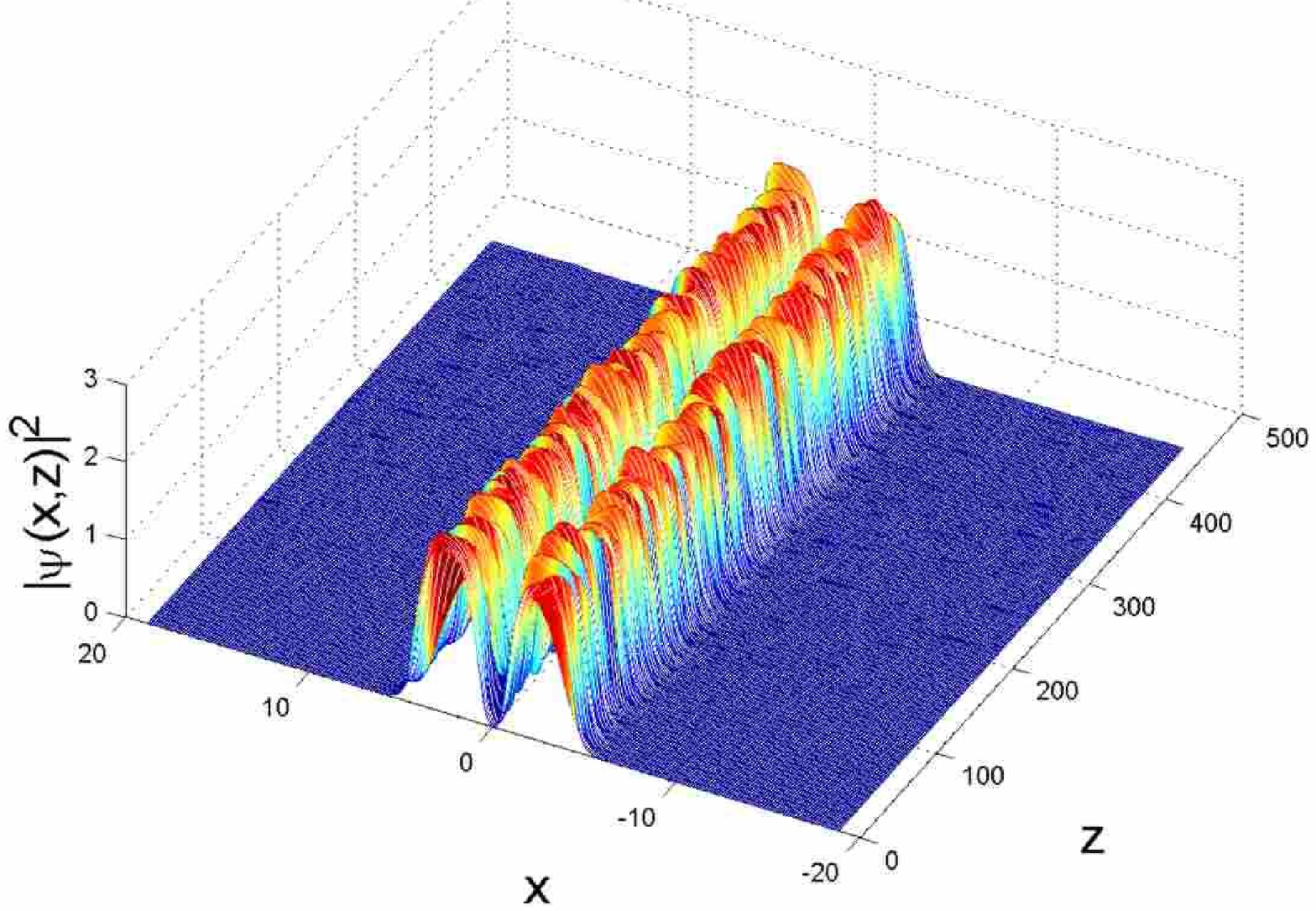}} \subfigure[]{%
\includegraphics[width=2.0in]{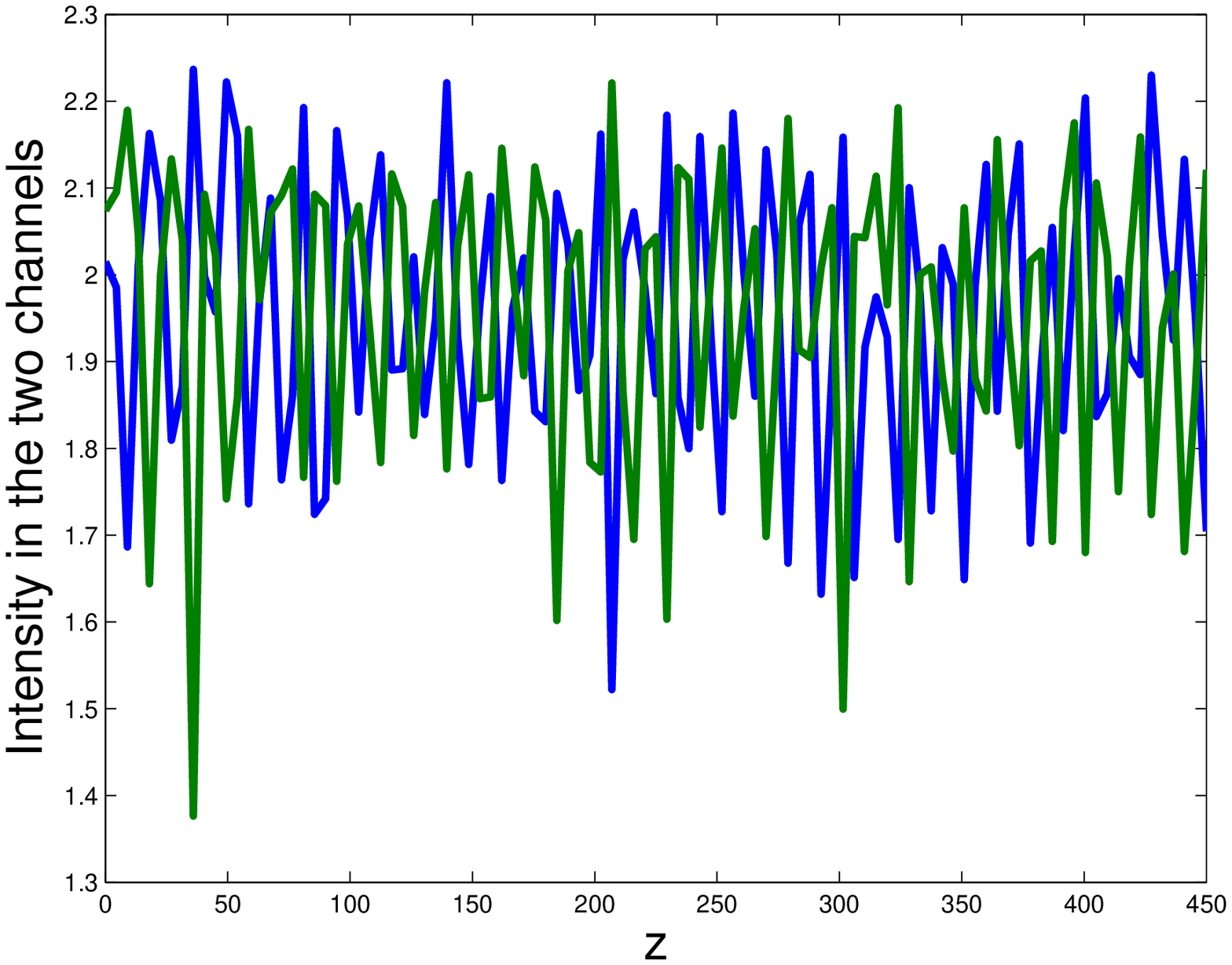}}
\caption{(Color online) The same as in Fig. \protect\ref{fig9}, with a
difference that the separation between the two channels is $L=2$ instead of $%
L=1$. The stronger coupling between the channels results in the formation of
a chaotic breather. The instability growth rate for the underlying
stationary soliton is $\protect\sigma \approx 0.5$.}
\label{fig10}
\end{figure}

\begin{figure}[tbp]
\includegraphics[width=3.0in]{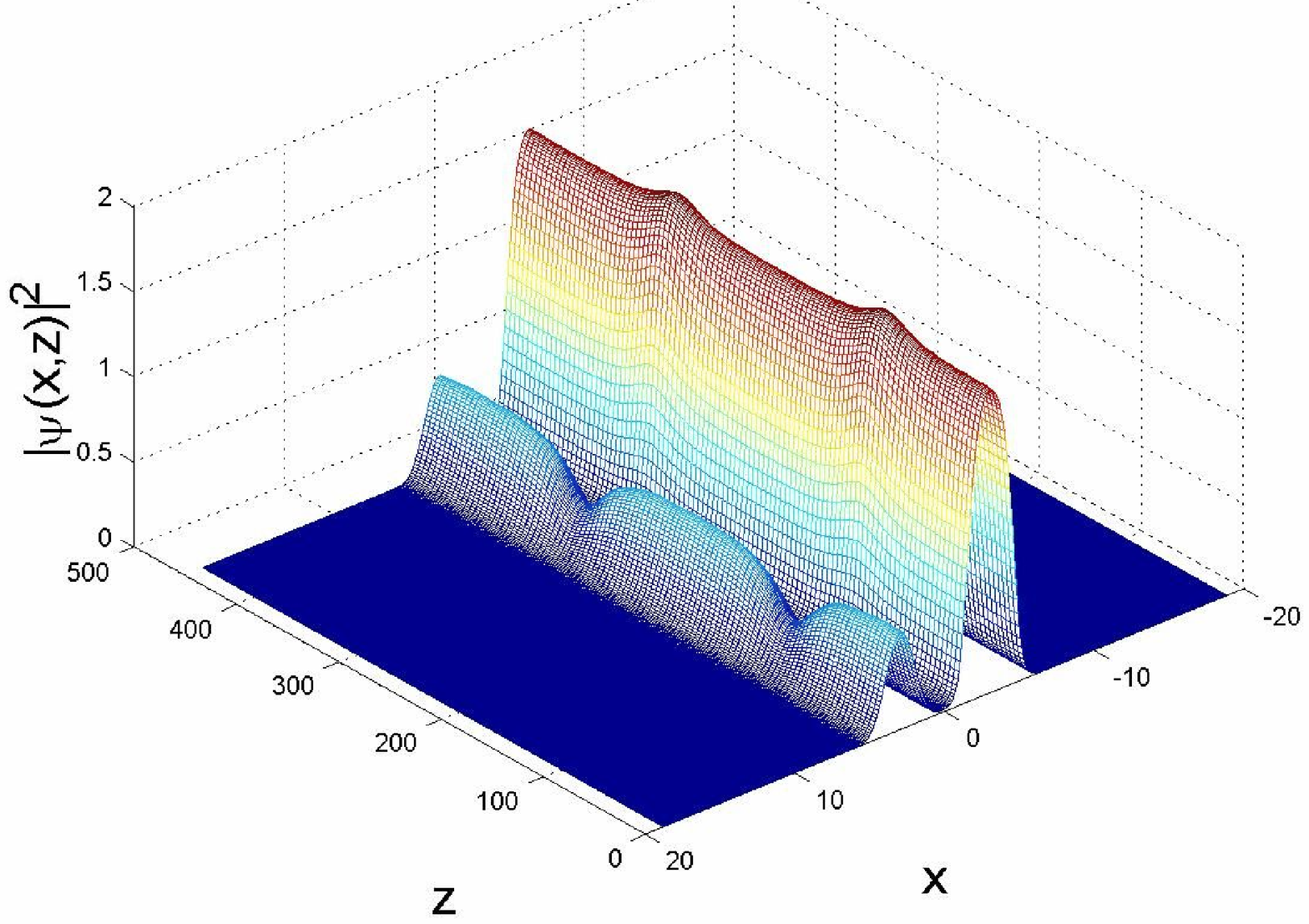}
\caption{(Color online) The same as in Fig. \protect\ref{fig8} for an
unstable unipolar composite state. Parameters are $V_{0}=3$, $D=4$, $L=2$,
and $k=2.85$. The instability growth rate for this state is $\protect\sigma %
\approx 0.05$.}
\label{fig11}
\end{figure}

Simulations of the perturbed evolution of the other three species, that
should be unstable, reveal that the instability occurs indeed (with the
exception of the above-mentioned stabilization of the broad antisymmetric
state in the case of very small $L$, through the stability exchange with the
narrow antisymmetric state). However, development of the instability does
not destroy the solitons, but rather transforms them into \emph{persistent
breathers}, as shown in Figs. \ref{fig8}-\ref{fig11}. This outcome of the
nonlinear evolution is observed despite the fact that the instability growth
rate for small perturbations is always real, while the transition to
breathers would be more straightforward in the case of an oscillatory
instability, accounted for by pairs of complex conjugate eigenvalues.
Accordingly, the period of the established oscillations of the breather
depends, although weakly, on the size of the initial perturbation which
destabilized the underlying stationary soliton (in the case shown in Fig. %
\ref{fig8}, the period will be larger by a factor $\simeq 1.5$ if the
amplitude of the initial perturbation is smaller by a factor of $10$).

The breathers formed from the unstable symmetric and antisymmetric solitons,
such as those displayed in Figs. \ref{fig8} and \ref{fig9}, feature not only
the phase shift of $\pi $ between the oscillations in the two channels, but
also SSB (spontaneous symmetry breaking), as clearly manifested, in panels
(b) of the figures, by the difference in the peak powers in the channels.
Additional simulations demonstrate that, unlike the period of the
oscillations, the size of the symmetry breaking, if measured as the
difference between the maximum values of the peak powers, practically does
not depend on the size of the initial perturbation.

In most cases, the breathers formed from unstable solitons exhibit strictly
periodic oscillations, as in Figs. \ref{fig8}(a) and \ref{fig11}. However,
one species of the unstable stationary states, \textit{viz}., broad
antisymmetric ones, may transform itself into a breather which features
(seemingly) \emph{chaotic} oscillations. Some irregularity is seen in the
evolution of the peak powers in Fig. \ref{fig9}(b). The increase of the
coupling between the channels, i.e., a decrease in $L$, makes the
oscillations manifestly chaotic, see Fig. \ref{fig10}. Note that the
instability growth rate of the unstable solitons, $\sigma $, which give rise
to chaotic breathers is much higher than in cases when periodic breathers
are generated: for instance, it is, respectively, $\sigma \approx 0.05$ and $%
\sigma \approx 0.5$ in Figs. \ref{fig8} and \ref{fig10}. On the other hand,
Fig. \ref{fig10}(b) suggests that the chaotic breather restores, on average,
the originally broken dynamical symmetry between the two channels. Indeed,
in that case the average values of the peak power in the channels are equal,
both being $1.955$.

\section{Bifurcation diagrams}
\label{sec4}

The global description of various stationary soliton families found in the
present model, and of links between them is provided by diagrams that show
the integral power of relevant solutions, $P$, as a function of the
propagation constant, $k$, along with diagrams that display an effective
asymmetry of the solutions versus $P$. The latter characteristic is defined
as follows:%
\begin{equation}
\varepsilon \equiv \frac{\int_{0}^{+\infty }\left\vert \psi (x)\right\vert
^{2}dx-\int_{-\infty }^{0}\left\vert \psi (x)\right\vert ^{2}dx}{P}
\label{epsilon}
\end{equation}%
[recall $P\equiv \int_{-\infty }^{0}\left\vert \psi (x)\right\vert
^{2}dx+\int_{0}^{+\infty }\left\vert \psi (x)\right\vert ^{2}dx$ is the
total power of the spatial soliton]. The form of the bifurcation diagrams
turns out to be quite different in the cases of strong and weak coupling
between the two channels, which correspond to small and large values of
thickness $L$ of the barrier between the channels. These two situations are
presented below separately, with labels identifying the eight basic types of
the solution branches in the plots as follows:

$\mathrm{A}$ and $\mathrm{B}$ -- symmetric stationary solutions of the broad
and narrow types, respectively;

$\mathrm{A}^{\prime }$ and $\mathrm{B}^{\prime }$ -- antisymmetric solutions
of the broad and narrow types, respectively;

$\mathrm{C}$ and $\mathrm{C}^{\prime }$ -- unipolar and bipolar composite
states, respectively;

$\mathrm{D}$ and $\mathrm{E}$ -- single-sided broad and narrow solutions,
respectively.

\subsection{Strong coupling}

\begin{figure}[tbp]
\subfigure[]{\includegraphics[width=1.75in]{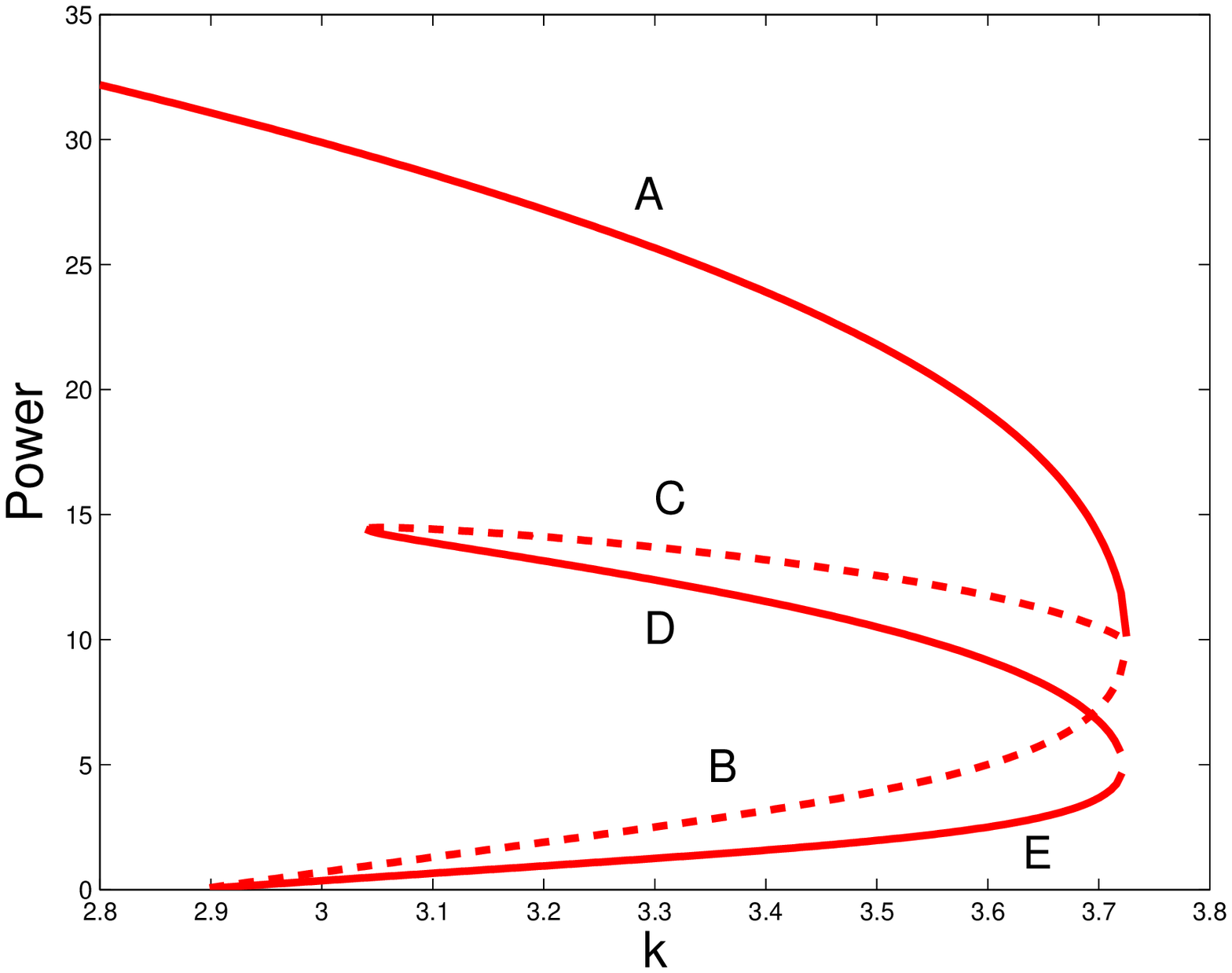}} \subfigure[]{%
\includegraphics[width=1.75in]{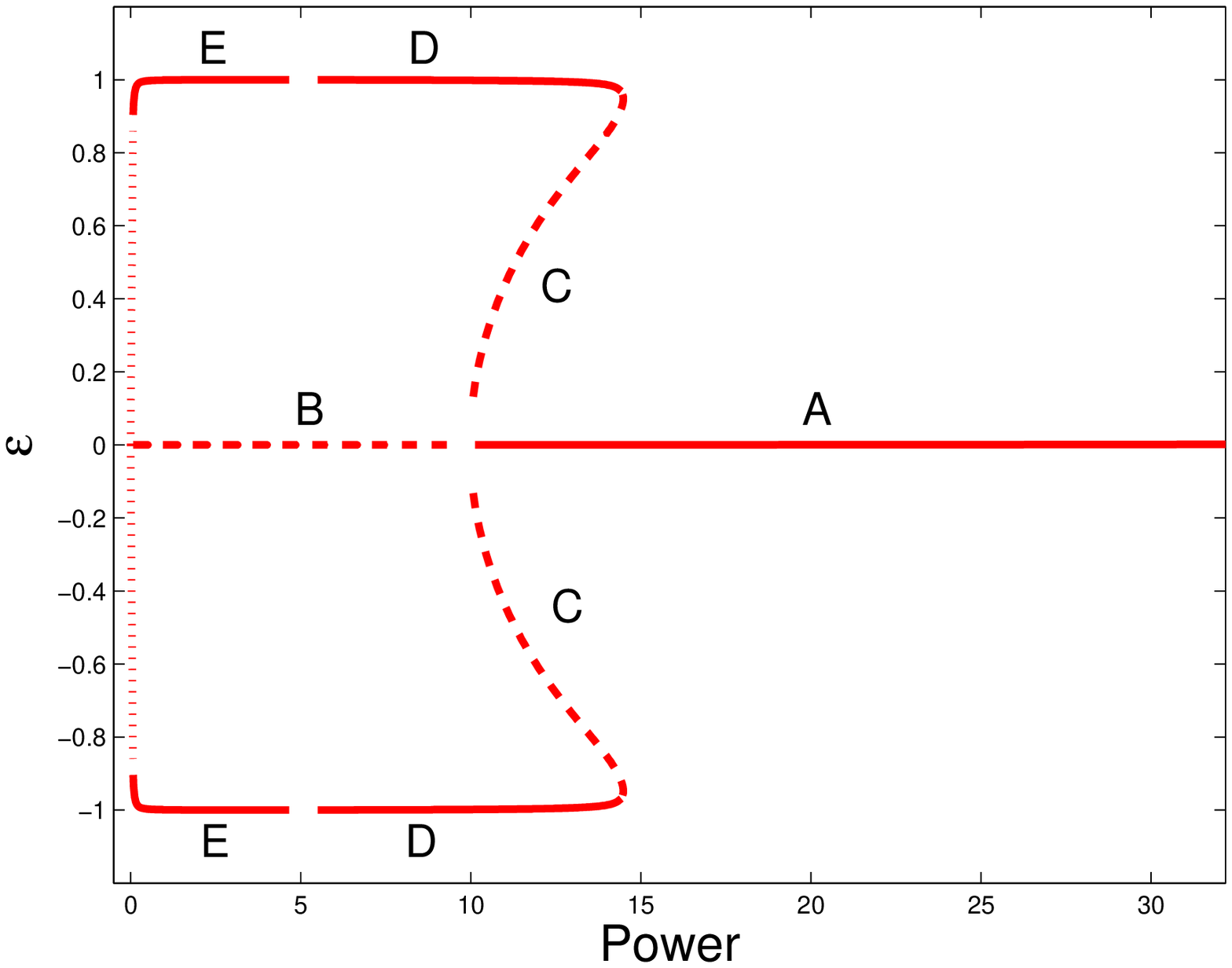}}
\caption{(Color online) Generic examples of the bifurcation diagrams in the
strongly-coupled system, for $V_{0}=3$, $D=8$, $L=1$. This set of the
diagrams displays symmetric, unipolar composite, and single-sided branches
of the spatial-soliton solutions. (a) The integral power ($P$) versus the
propagation constant; (b) the asymmetry parameter (defined as per Eq. (%
\protect\ref{epsilon})) versus $P$. Labels A, B, C, D and E are
defined in the text.} \label{fig12}
\end{figure}

\begin{figure}[tbp]
\subfigure[]{\includegraphics[width=1.75in]{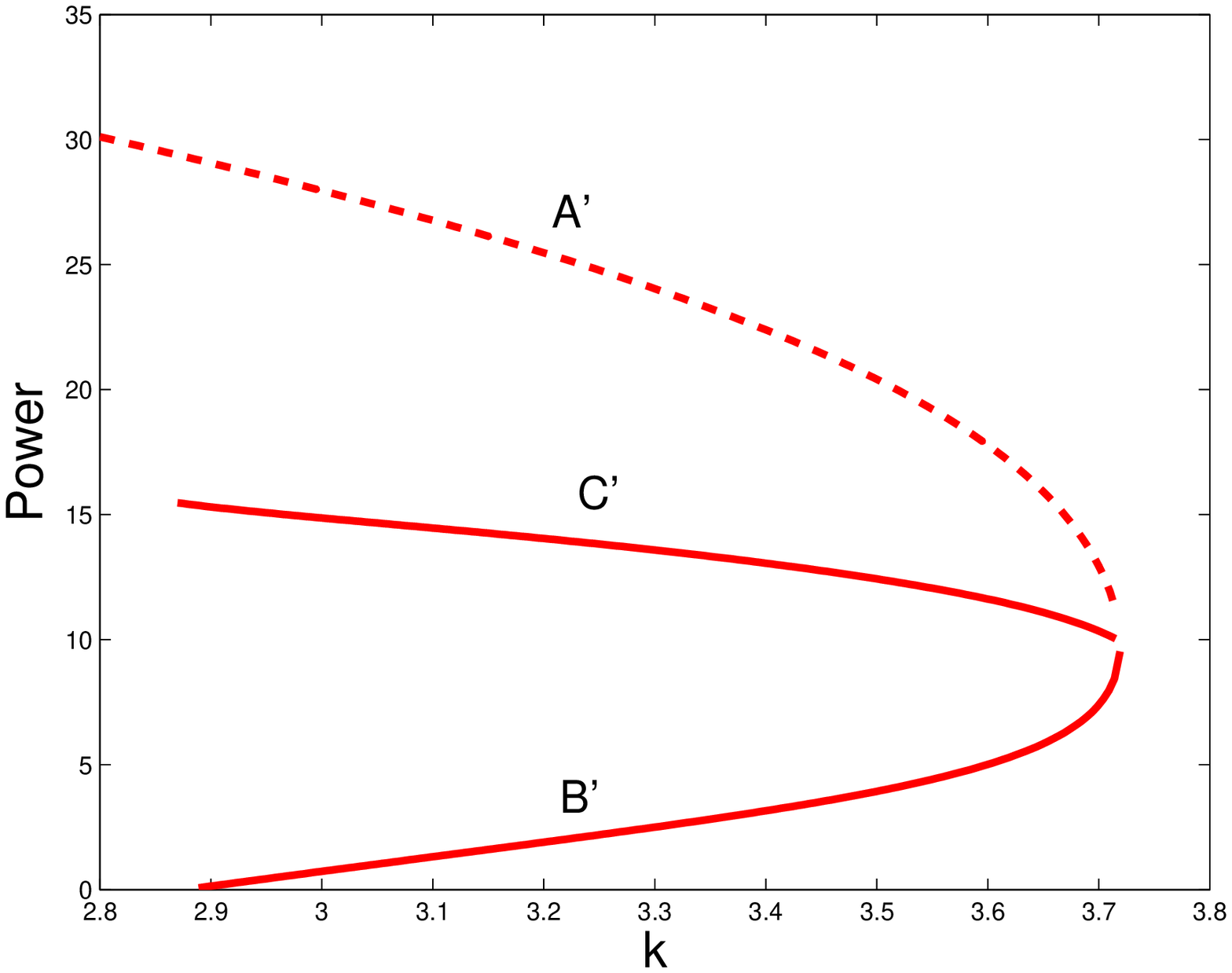}} \subfigure[]{%
\includegraphics[width=1.75in]{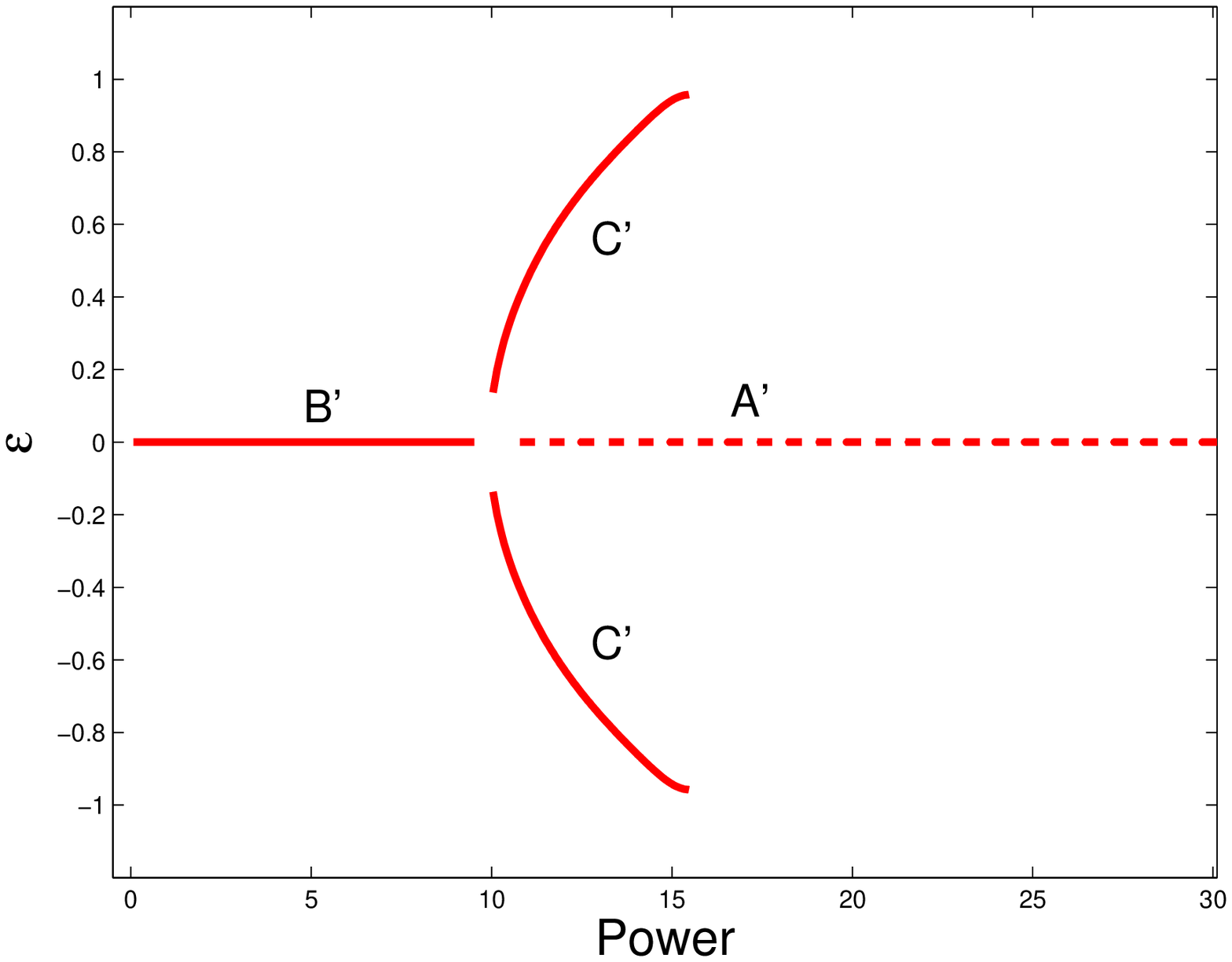}}
\caption{(Color online) The same as in Fig. \protect\ref{fig12}, but for
branches of the antisymmetric and bipolar composite states. Labels A$%
^{\prime }$, B$^{\prime }$, and C$^{\prime }$ are defined in the text.}
\label{fig13}
\end{figure}

Generic bifurcation diagrams for the system with a relatively strong
interaction between the beams trapped in the two channels are displayed in
Figs. \ref{fig12} and \ref{fig13}. Following the usual convention,
continuous and dashed curves in these plots represent stable and unstable
solution branches. Discontinuities at turning points in panels \ref{fig12}%
(a) and \ref{fig13}(a) are due to the (well-known) problem with poor
convergence of numerical solutions in a vicinity of such points. For
the same reason, the left, nearly vertical, segments of branch
$\mathrm{E}$ in Figs. \ref{fig12}(b) and \ref{fig13}(b) (and also in
Figs. \ref{fig14}(b) and \ref{fig15}(b) below) were found with a low
accuracy, therefore they are shown by dots. Note that values of $k$
at the two right turning points in Figs. \ref{fig12}(a) and
\ref{fig13}(a), as well as in Figs. \ref{fig14}(a) and
\ref{fig15}(a) below, are close but not equal.

We stress that the bifurcations at points where branches $\mathrm{A}$, $%
\mathrm{B}$ and $\mathrm{C}$ meet in Fig. \ref{fig12}(a), as well as where $%
\mathrm{A}^{\prime }$, $\mathrm{B}^{\prime }$ and $\mathrm{C}^{\prime }$
meet in Fig. \ref{fig13}(a), actually involve not three but four different
branches (in compliance with general principles of the bifurcation theory
\cite{JosephIooss}). Indeed, Figs. \ref{fig12}(b) and \ref{fig13}(b) clearly
demonstrate that branches $\mathrm{C}$ and $\mathrm{C}^{\prime }$ exist each
in two copies, that are mirror images to each other: one is composed of a
narrow soliton in the left channel and a broad one in the right channel, and
vice versa. It is noteworthy that branches $\mathrm{C}$ and $\mathrm{D}$,
which represent, severally, the unstable unipolar composite solitons and
stable single-sided broad solitons, meet and disappear through a tangent
(alias saddle-node \cite{JosephIooss}) bifurcation in Fig. \ref{fig10}(a).
Also worthy to note is the fact that branches $\mathrm{D}$ and $\mathrm{E}$,
which meet and disappear at the lower left turning point in Fig. \ref{fig10}%
(a), are both stable, contrary to the ``naive" expectation that one of them
ought to be unstable (this fact was already discovered in the single-channel
CQ model \cite{we}).

The bifurcation diagram in the plane of $P$ and $\varepsilon $ actually
describes the SSB, in terms of stationary solutions. Note that the diagram
in Fig. \ref{fig12}(b) features a loop, which is a characteristic feature of
the SSB in models with saturable \cite{Canberra} and CQ nonlinearities. In
the latter context, closed bifurcation loops were found in the discrete
(lattice) version of the CQ model \cite{Ricardo}, and in a system of two
linearly coupled NLS equations with the CQ nonlinearity \cite{Lior}. The
presence of the bifurcation loop implies that the broken symmetry is
eventually restored, under the action of the quintic term. On the other
hand, the diagram in the same plane for solution branches $\mathrm{A}%
^{\prime }$, $\mathrm{B}^{\prime }$ and $\mathrm{C}^{\prime }$ in Fig. \ref%
{fig13}(b) does not display a closed loop. In fact, two mutually
symmetric branches $\mathrm{C}^{\prime }$ in this diagram terminate
because of difficulties with the numerical continuation to larger
values of $P$; on the other hand, the branch of broad antisymmetric
solitons, $\mathrm{A}^{\prime } $, can be continued easily, and the
continuation does not reveal its stabilization at any particular
value of $P$ (in direct simulations, the broad antisymmetric soliton
always turns into a persistent breather, cf. Fig. \ref{fig8}(b)),
hence it is plausible that the configuration observed in Fig.
\ref{fig11}(b) is not going to form a closed bifurcation loop. As
concerns the continuation of branch $\mathrm{A}$ in Fig.
\ref{fig12}(b) (and also in Fig. \ref{fig14}(b) below) to
$P\rightarrow \infty $, this happens with $k\rightarrow k_{\max
}\equiv 3/4$, when state $\mathrm{A}$ becomes
asymptotically similar to CW state $\psi _{\mathrm{CW}}^{(+)}$, see Eq. (\ref%
{CW}). Similarly, state $\mathrm{A}^{\prime }$ in Figs. \ref{fig13}(b), and
also in \ref{fig15}(b) below, become similar, in the same limit to a dark
soliton in the free space.

\subsection{Weak coupling}

\begin{figure}[tbp]
\subfigure[]{\includegraphics[width=1.75in]{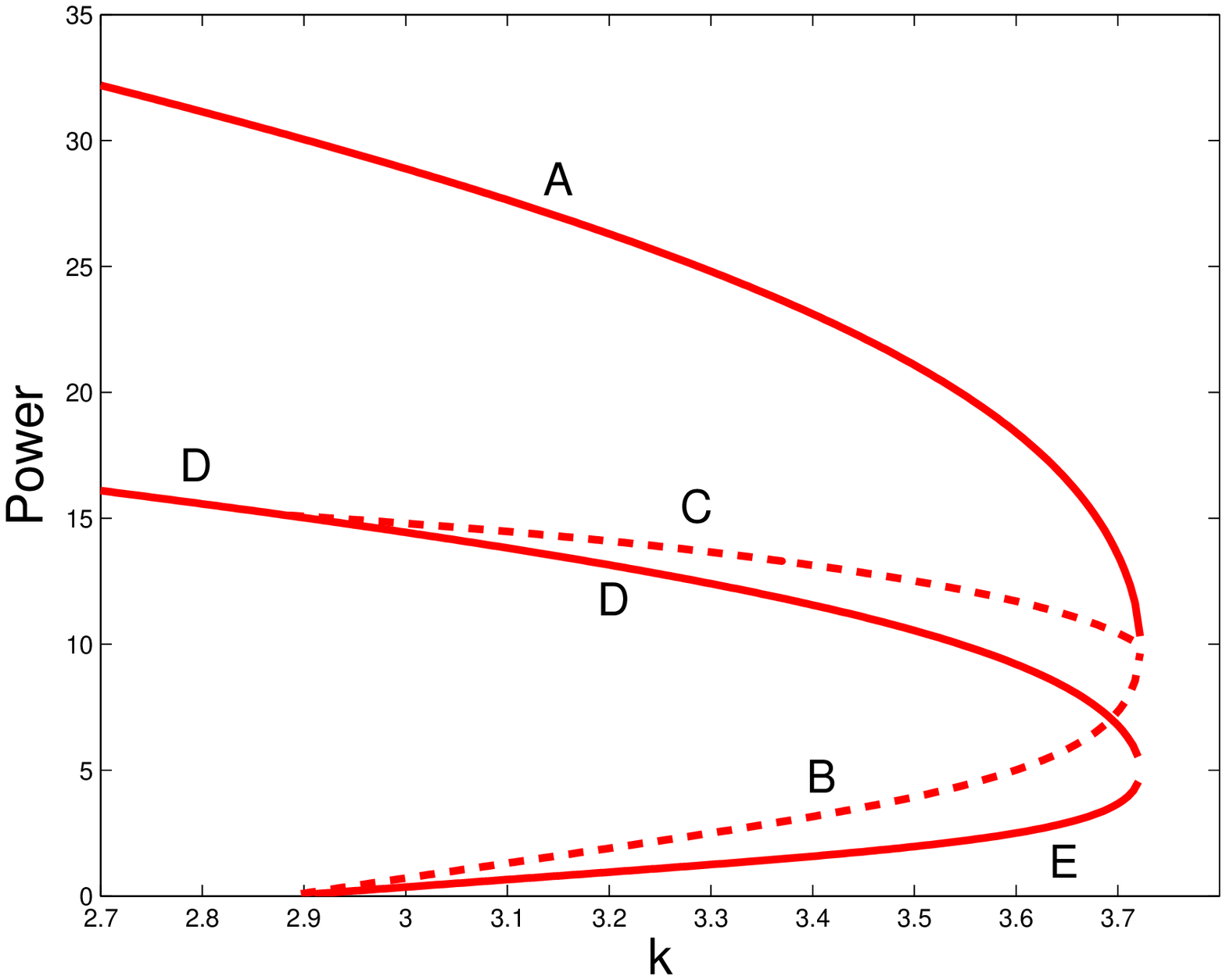}} \subfigure[]{%
\includegraphics[width=1.75in]{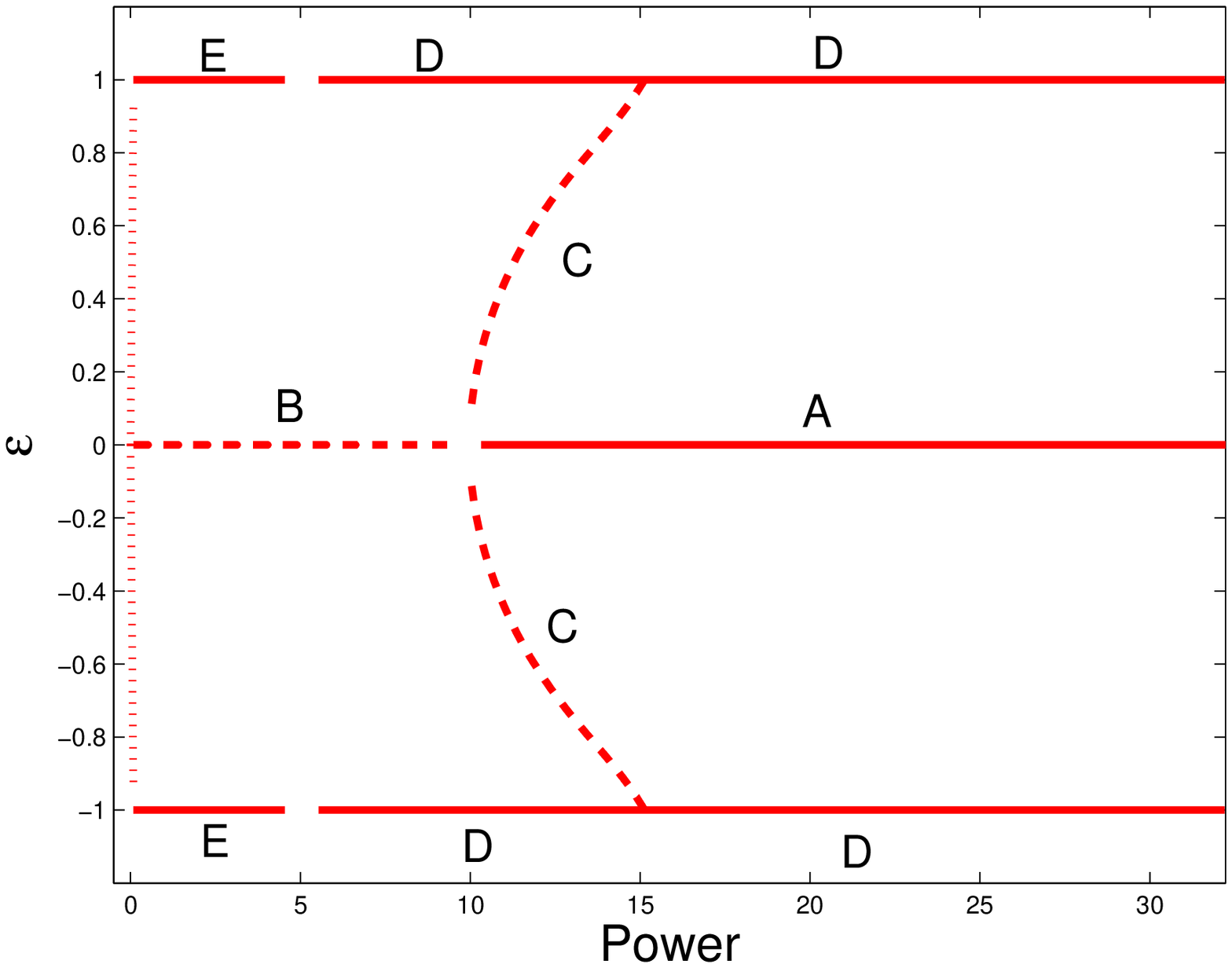}}
\caption{(Color online) The same as in Fig. \protect\ref{fig12}, but for the
weakly-coupled system, with $V_{0}=3$, $D=8$, and $L=8$.}
\label{fig14}
\end{figure}

\begin{figure}[tbp]
\subfigure[]{\includegraphics[width=1.75in]{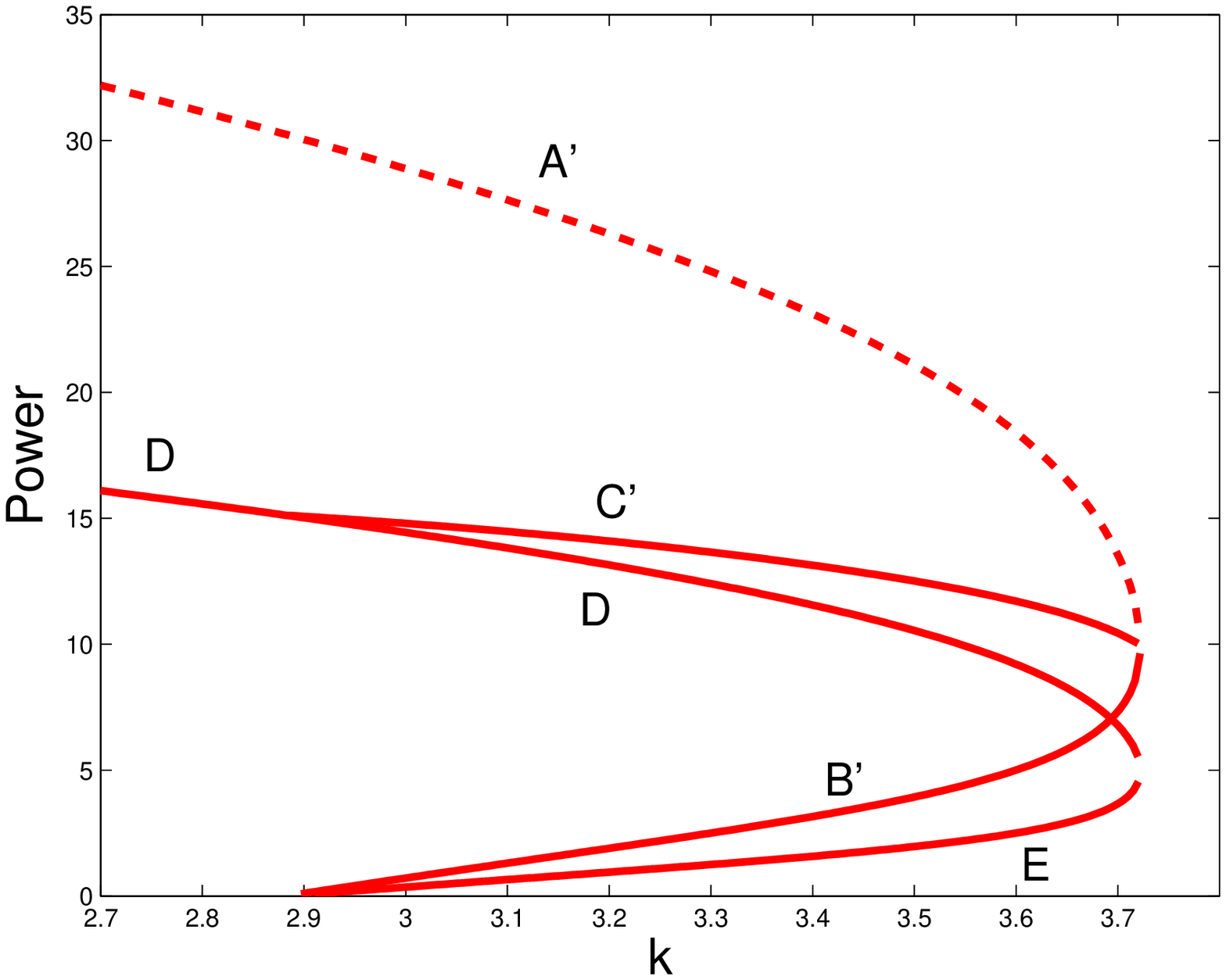}} \subfigure[]{%
\includegraphics[width=1.75in]{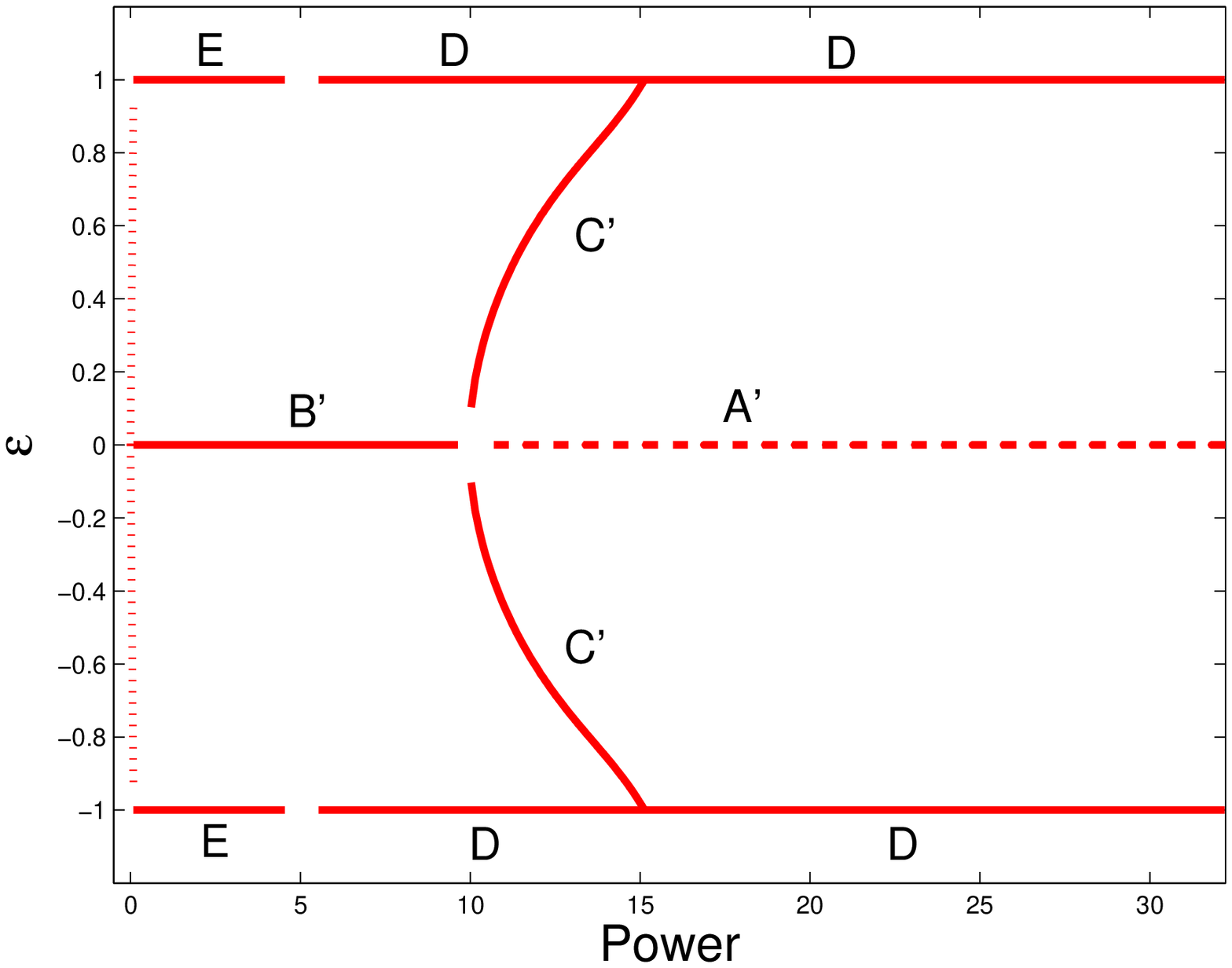}}
\caption{(Color online) The same as in Fig. \protect\ref{fig13}, with
parameters identical to those in Fig. \protect\ref{fig14}, but also
including branches $\mathrm{D}$ and $\mathrm{E}$ of the single-sided
solitons.}
\label{fig15}
\end{figure}

At essentially larger values of the thickness of the buffer layer between
the channels ($L$), which implies weak interaction between them, the
character of the bifurcation diagrams changes qualitatively, as shown in
Figs. \ref{fig14} and \ref{fig15}. In this situation, the tangent
bifurcation through which branches $\mathrm{C}$ and $\mathrm{D}$ disappeared
in the case of the strong inter-channel coupling does not take place, cf.
Fig. \ref{fig12}(a). Instead, there occurs another bifurcation that involves
solution families $\mathrm{C}$, $\mathrm{C}^{\prime }$ and $\mathrm{D}$ (in
fact, family $\mathrm{D}$ of the stable broad single-sided solutions
consists of two branches, one to the right of the bifurcation point, and the
continuation to the left of it). This bifurcation manifests itself in the
diagrams that involve the sets of both symmetric and antisymmetric states,
i.e., $\left( \mathrm{A},\mathrm{B},\mathrm{C}\right) $ in Fig. \ref{fig14}%
(a), and $\left( \mathrm{A}^{\prime },\mathrm{B}^{\prime },\mathrm{C}%
^{\prime }\right) $ in Fig. \ref{fig15}(a).

In the plane of $\left( P,\varepsilon \right) $, the continuation of
branches $\mathrm{D}$ toward larger values of the power does not reveal any
trend to forming a closed loop. Instead, the branches remain virtually
identical to lines $\varepsilon =\pm 1$, which is obvious in Figs. \ref%
{fig14}(b) and \ref{fig15}(b). This feature is explained by the propensity
of the broad single-sided soliton to turn into the front state (which
obviously has $\varepsilon =\pm 1$), as shown above in Fig. \ref{fig5}.

In both cases of the strong and weak coupling, the model gives rise to four
bifurcations: those involving sets of branches $\left( \mathrm{A},\mathrm{B},%
\mathrm{C}\right) $, $\left( \mathrm{A}^{\prime },\mathrm{B}^{\prime },%
\mathrm{C}^{\prime }\right) $, $\left( \mathrm{C},\mathrm{D}\right) $ and $%
\left( \mathrm{D},\mathrm{E}\right) $ in Figs. \ref{fig12} and \ref{fig13},
and sets $\left( \mathrm{A},\mathrm{B},\mathrm{C}\right) $, $\left( \mathrm{A%
}^{\prime },\mathrm{B}^{\prime },\mathrm{C}^{\prime }\right) $, $\left(
\mathrm{C},\mathrm{C}^{\prime },\mathrm{D}\right) $ and $\left( \mathrm{D},%
\mathrm{E}\right) $ in Figs. \ref{fig14} and \ref{fig15}. The transition
between the bifurcation diagrams corresponding to these two cases amounts to
the well-known generic type of the rearrangement, when a branch splits off
from a pitchfork bifurcation pattern, leaving behind a bifurcation of the
tangent type (in the present situation, the detaching branch is $\mathrm{C}%
^{\prime }$, as suggested by Fig. \ref{fig11}(a)). We did not aim to
find an exact value of $L$ (for fixed $V_{0}$ and $D$) at which the
bifurcation changes its character.

\section{Conclusion}
\label{sec5}

The aim of this work was to explore spatial solitons that can be
trapped in a planar waveguide which combines two fundamental
ingredients, namely, the dual-channel configuration and competing
(cubic-quintic) nonlinearity. We have found eight species of the
solitons, including symmetric and antisymmetric ones of the broad
and narrow types, unipolar and bipolar composite solitons, and the
broad and narrow species of single-sided solitons. In contrast to
this situation, previously studied models combining two-channel
traps with non-competing nonlinearities (of the cubic and saturable
types) gave rise to three species of the states (symmetric,
antisymmetric, and single-sided asymmetric ones). The stability of
the eight families was explored through the computation of
perturbation eigenvalues and by means of direct simulations. It has
been concluded that three families are unstable, \textit{viz}.,
narrow symmetric, broad antisymmetric, and unipolar composite
states, each of the other five species being stable as a whole. All
instabilities against infinitesimal perturbations are accounted for
by real eigenvalues. However, unstable solitons are not destroyed by
the instability, but rather turn into robust breathers, which
feature dynamical SSB (spontaneousness symmetry breaking), between
the field intensities in the two channels. The breathers generated
by strongly unstable broad antisymmetric solitons turn out to be
chaotic (rather than periodic). It is interesting that the broad and
narrow spatial solitons of the antisymmetric type switch their
stability in the limit when the two-channel setting is going to
merge into the single-channel one: the broad state becomes stable,
while the narrow one is destabilized. Four bifurcations linking
different branches of the stationary states have been found. The
bifurcation diagrams may be of two different types, corresponding to
the strong or weak coupling between the two channels. In the latter
case, the bifurcation diagram involving the symmetric and unipolar
composite states features a closed loop (as seen in Fig.
\ref{fig12}). An additional finding, which is relevant to the
single-channel setting as well, is the stable front-shaped state (as
shown in Fig. \ref{fig5}).

Numerous stable spatial solitons possible in this model may find
applications to all-optical data-processing schemes. Moreover, the potential
of the two-channel model is not exhausted by the analysis of the eight types
of the trapped states reported in this work. Additional species of spatial
soliton are possible in it, the form of symmetric, antisymmetric, composite
and single-sided configurations based on stable broad dipoles (intrinsically
antisymmetric states) trapped in each channel, as well as complexes
involving a fundamental soliton in one channel and a broad dipole in the
other. A challenging generalization is also possible for a two-dimensional
model with the CQ nonlinearity and two parallel trapping channels. In that
case, each channel may carry a fundamental two-dimensional soliton or a
vortex.

\section*{Acknowledgement}

This work was supported, in a part, by Israel Science Foundation through
Center-of-Excellence grant No. 8006/03.

\end{document}